\documentclass[11pt]{article}

\usepackage{latexsym}
\usepackage{amssymb}
\usepackage{epsfig}
\usepackage{amsfonts}
\usepackage{color}
\usepackage{amsmath}
\begin{document}

\newcommand{\bull}{\rule{.85ex}{1ex} \par \bigskip}
\newenvironment{sketch}{\noindent {\bf Proof (sketch):\ }}{\hfill \bull}
\newenvironment{example}{\begin{exmp} \rm }{\hfill $\Box$ \end{exmp}}

\newtheorem{theorem}{Theorem}[section]
\newtheorem{definition}[theorem]{Definition}
\newtheorem{proposition}[theorem]{Proposition}
\newtheorem{lemma}[theorem]{Lemma}
\newtheorem{corollary}[theorem]{Corollary}
\newtheorem{conjecture}[theorem]{Conjecture}
\newtheorem{exmp}[theorem]{Example}
\newtheorem{notation}[theorem]{Notation}
\newtheorem{problem}{Problem}
\newtheorem{remark}[theorem]{Remark}
\newtheorem{observation}[theorem]{Observation}

\newcommand{\QCSP}[1]{\mbox{\rm QCSP$(#1)$}}
\newcommand{\CSP}[1]{\mbox{\rm CSP$(#1)$}}
\newcommand{\MCSP}[1]{\mbox{{\sc Max CSP}$(#1)$}}
\newcommand{\wMCSP}[1]{\mbox{\rm weighted Max CSP$(#1)$}}
\newcommand{\cMCSP}[1]{\mbox{\rm cw-Max CSP$(#1)$}}
\newcommand{\tMCSP}[1]{\mbox{\rm tw-Max CSP$(#1)$}}
\renewcommand{\P}{\mbox{\bf P}}
\newcommand{\G}[1]{\mbox{\rm I$(#1)$}}
\newcommand{\NE}[1]{\mbox{$\neq_{#1}$}}

\newcommand{\NP}{\mbox{\bf NP}}
\newcommand{\NL}{\mbox{\bf NL}}
\newcommand{\PO}{\mbox{\bf PO}}
\newcommand{\NPO}{\mbox{\bf NPO}}
\newcommand{\APX}{\mbox{\bf APX}}

\newcommand{\GIF}[3]{\ensuremath{h_\{{#2},{#3}\}^{#1}}}

\newcommand{\Spmod}{\mbox{\rm Spmod}}
\newcommand{\Sbmod}{\mbox{\rm Sbmod}}

\newcommand{\Inv}[1]{\mbox{\rm Inv($#1$)}}
\newcommand{\Pol}[1]{\mbox{\rm Pol($#1$)}}
\newcommand{\sPol}[1]{\mbox{\rm s-Pol($#1$)}}

\newcommand{\un}{\underline}
\newcommand{\ov}{\overline}
\def\ar{\hbox{ar}}
\def\vect#1#2{#1 _1\zdots #1 _{#2}}
\def\zd{,\ldots,}
\let\sse=\subseteq
\let\la=\langle
\def\lla{\langle\langle}
\let\ra=\rangle
\def\rra{\rangle\rangle}
\let\vr=\varrho
\def\vct#1#2{#1 _1\zd #1 _{#2}}
\newcommand{\va}{{\bf a}}
\newcommand{\vb}{{\bf b}}
\newcommand{\vc}{{\bf c}}
\newcommand{\ba}{{\bf a}}
\newcommand{\bx}{{\bf x}}
\newcommand{\by}{{\bf y}}
\def\Z{{\bur Z^+}}
\def\R{{\bur R}}
\def\D{{\cal D}}
\def\F{{\cal F}}
\def\I{{\cal I}}
\def\C{{\cal C}}
\def\K{{\cal K}}
\def\Q{{\cal Q}}
\def\U{{\cal U}}
\def\Lat{{\cal L}}

\def\2mat#1#2#3#4#5#6#7#8{
\begin{array}{c|cc}
$~$ & #3 & #4\\
\hline
#1 & #5& #6\\
#2 & #7 & #8 \end{array}}

\font\tenbur=msbm10
\font\eightbur=msbm8
\def\bur{\fam11}
\textfont11=\tenbur \scriptfont11=\eightbur \scriptscriptfont11=\eightbur
\font\twelvebur=msbm10 scaled 1200
\textfont13=\twelvebur \scriptfont13=\tenbur \scriptscriptfont13=\eightbur




\newcounter{A}
\newcounter{prgline} 


\newcommand{\citelist}[1]{\raisebox{.2ex}{[}#1\raisebox{.2ex}{]}}
\newcommand{\scite}[1]{\citeauthor{#1}, \citeyear{#1}}
\newcommand{\shortcite}[1]{\cite{#1}}
\newcommand{\mciteii}[2]{\citeauthor{#1}, \citeyear{#1}, %
\citeyear{#2}}
\newcommand{\mciteiii}[3]{\citeauthor{#1}, \citeyear{#1}, %
\citeyear{#2}, \citeyear{#3}}

\newcommand{\multiciteii}[2]{\citelist{\scite{#1}, \citeyear{#2}}}
\newcommand{\multiciteiii}[3]%
  {\citelist{\scite{#1}, \citeyear{#2}, \citeyear{#3}}}


\renewcommand{\phi}{\varphi}
\renewcommand{\epsilon}{\varepsilon}

\newcommand{\draft}{\begin{center}\huge Draft!!! \end{center}}
\newcommand{\void}{\makebox[0mm]{}}     


\renewcommand{\text}[1]{\mbox{\rm \,#1\,}}        

\newcommand{\tand}{\text{\ and\ }}
\newcommand{\tor}{\text{\ or\ }}
\newcommand{\tif}{\text{\ if\ }}
\newcommand{\tiff}{\text{\ iff\ }}
\newcommand{\tfor}{\text{\ for\ }}
\newcommand{\tforall}{\text{\ for all\ }}
\newcommand{\totherwise}{\text{\ otherwise}}


\newcommand{\fnl}{\void \\}             

\newcommand{\pushlist}[1]{\setcounter{#1}{\value{enumi}} \end{enumerate}}
\newcommand{\poplist}[1]{\begin{enumerate} \setcounter{enumi}{\value{#1}}}


\renewcommand{\emptyset}{\varnothing}  
\newcommand{\union}{\cup}               
\newcommand{\intersect}{\cap}           
\newcommand{\setdiff}{-}                
\newcommand{\compl}[1]{\overline{#1}}   
\newcommand{\card}[1]{{|#1|}}           
\newcommand{\set}[1]{\{{#1}\}} 
\newcommand{\st}{\ |\ }                 
\newcommand{\suchthat}{\st}             
\newcommand{\cprod}{\times}             
\newcommand{\powerset}[1]{{\bf 2}^{#1}} 

\newcommand{\tuple}[1]{\langle{#1}\rangle}  
\newcommand{\seq}[1]{\langle #1 \rangle}
\newcommand{\emptyseq}{\seq{}}
\newcommand{\floor}[1]{\left\lfloor{#1}\right\rfloor}
\newcommand{\ceiling}[1]{\left\lceil{#1}\right\rceil}

\newcommand{\map}{\rightarrow}
\newcommand{\fncomp}{\!\circ\!}         

\newcommand{\transclos}[1]{#1^+}
\newcommand{\reduction}[1]{#1^-}        

\newtheorem{defnx}{Definition}
\newtheorem{axiomx}[defnx]{Axiom}
\newtheorem{theoremx}[defnx]{Theorem}
\newtheorem{propositionx}[defnx]{Proposition}
\newtheorem{lemmax}[defnx]{Lemma}
\newtheorem{corollx}[defnx]{Corollary}
\newtheorem{algx}[defnx]{Algorithm}
\newtheorem{exx}[defnx]{Example}
\newtheorem{factx}[defnx]{Fact}

\newcommand{\QED}{\nopagebreak[4]{\makebox[1mm]{}\hfill$\Box$}}
\newenvironment{proof}{\noindent {\bf \noindent Proof: }}{\QED \\}
\newenvironment{proofsk}{\noindent {\bf \noindent Proof sketch: }}{\QED
\vspace{-\baselineskip}}


\newlength{\prgindent}
\newenvironment{program}{
  \begin{list}%
   {\arabic{prgline}}%
   {
   \usecounter{prgline}
   \setlength{\prgindent}{0em}
   \setlength{\parsep}{0em}
   \setlength{\itemsep}{0em}
   \setlength{\labelwidth}{1em}
   \setlength{\labelsep}{1em}
   \setlength{\leftmargin}{\labelwidth}
   \addtolength{\leftmargin}{\labelsep}
   \setlength{\topsep}{0em}
   \setlength{\parskip}{0em} } }%
 {\end{list}}

\newif\ifprgendtext
\prgendtexttrue

\newenvironment{prgblock}{\addtolength{\prgindent}{\labelsep}}%
{\addtolength{\prgindent}{-\labelsep}}
\newcommand{\prgbeginblock}{\addtolength{\prgindent}{\labelsep}}
\newcommand{\prgendblock}{\addtolength{\prgindent}{-\labelsep}}
\newcommand{\prgcndendblock}[1]{\addtolength{\prgindent}{-\labelsep}
 \ifprgendtext \prglin\prgres{#1}\fi}

\newcommand{\prglin}{\rm \item\hspace{\prgindent}}
\newcommand{\prgcontlin}{\\  \hspace{\prgindent}}

\newcommand{\prgres}[1]{{\bf #1}}
\newcommand{\prgassn}{\leftarrow}
\newcommand{\prgname}[1]{{\it #1}}

\newcommand{\prgbegin}{\prglin\prgres{begin}\prgbeginblock}
\newcommand{\prgend}{\prgendblock\prglin\prgres{end\ }}
\newcommand{\prgnoend}{\prgendblock}

\newcommand{\prgif}{\prglin\prgres{if\ }}
\newcommand{\prgthen}{\prgres{\ then\ }\prgbeginblock}
\newcommand{\prgelse}{\prgendblock\prglin\prgres{else}\prgbeginblock}
\newcommand{\prgelsif}{\prgendblock\prglin\prgres{elsif\ }}
\newcommand{\prgendif}{\prgcndendblock{end if}}

\newcommand{\prgwhile}{\prglin\prgres{while\ }}
\newcommand{\prgfor}{\prglin\prgres{for\ }}
\newcommand{\prgdo}{\prgres{\ do}\prgbeginblock}
\newcommand{\prgrepeat}{\prgres{\ repeat}\prgbeginblock}
\newcommand{\prgloop}{\prglin\prgres{loop}\prgbeginblock}
\newcommand{\prgendloop}{\prgendblock\prglin\prgres{end loop}}
\newcommand{\prgendwhile}{\prgcndendblock{end while}}
\newcommand{\prgendfor}{\prgcndendblock{end for}}
\newcommand{\prguntil}{\prgendblock\prglin\prgres{until\ }}

\newcommand{\prgcomment}{\prglin\prgres{comment\ }\it }
\newcommand{\prgprocedure}{\prglin\prgres{procedure\ }}
\newcommand{\prgnil}{\prgres{\ nil}}
\newcommand{\prgtrue}{\prgres{\ true}}
\newcommand{\prgfalse}{\prgres{\ false}}
\newcommand{\prgnot}{\prgres{\ not\ }}
\newcommand{\prgand}{\prgres{\ and\ }}
\newcommand{\prgor}{\prgres{\ or \ }}
\newcommand{\prgfail}{\prgres{fail}}
\newcommand{\prgreturn}{\prgres{return\ }}
\newcommand{\prgaccept}{\prgres{accept}}
\newcommand{\prgreject}{\prgres{reject}}

\prgendtextfalse

\newcommand{\note}[1]{{\tt #1}}


\newcommand{\ie}{{\em ie.}}                
\newcommand{\eg}{{\em eg.}}
\newcommand{\paper}{paper}                

\newcommand{\emdef}{\em}                   
\newcommand{\rinterpretation}{${\Bbb R}$-interpretation}
\newcommand{\rmodel}{${\Bbb R}$-model}
\newcommand{\transp}{^{\rm T}}

\newcommand{\unprint}[1]{}
\newcommand{\blankline}{$\:$}

\newcommand{\Solv}{{\it TSolve}}
\newcommand{\Neg}{{\it Neg}}
\newcommand{\logname}{XX}

\newcommand{\props}{{\it props}}
\newcommand{\rels}{{\it rels}}
\newcommand{\deduce}{\vdash_p}

\newcommand{\pform}{{\rm Pr}}
\newcommand{\axform}{{\rm AX}}
\newcommand{\axset}{{\bf AX}}
\newcommand{\resdeduce}{\vdash_{\rm R}}
\newcommand{\resaxdeduce}{\vdash_{\rm R,A}}

\newcommand{\cmis}{{\em \#mis}}
\newcommand{\combine}{{\em comb}}

\newcommand{\xcsp}{{\sc X-Csp}}
\newcommand{\csp}{{\sc Csp}}

\author{
Peter Jonsson, Mikael Klasson\\
Department of Computer and Information Science\\
University of Link\"oping, Sweden\\
$\texttt{peter.jonsson@ida.liu.se, mikkl@ida.liu.se}$\\
\and
Andrei Krokhin\\
Department of Computer Science\\
University of Durham, UK\\
$\texttt{andrei.krokhin@durham.ac.uk}$\\
}
\title{The approximability of three-valued {\sc Max CSP}}

\date{}
\maketitle
\bibliographystyle{plain}


\begin{abstract}
In the maximum constraint satisfaction problem ({\sc Max CSP}),
one is given a finite collection of (possibly weighted)
constraints on overlapping sets of variables, and the goal is to
assign values from a given domain to the variables so as to
maximize the number (or the total weight, for the weighted case)
of satisfied constraints. This problem is \NP-hard in general,
and, therefore, it is natural to study how restricting the allowed
types of constraints affects the approximability of the problem.
It is known that every Boolean (that is, two-valued) {\sc Max CSP}
problem with a finite set of allowed constraint types is either
solvable exactly in polynomial time or else \APX-complete (and
hence can have no polynomial time approximation scheme unless
$\P=\NP$). It has been an open problem for several years whether
this result can be extended to non-Boolean {\sc Max CSP}, which is
much more difficult to analyze than the Boolean case. In this
paper, we make the first step in this direction by establishing
this result for {\sc Max CSP} over a three-element domain.
Moreover, we present a simple description of all polynomial-time
solvable cases of our problem. This description uses the
well-known algebraic combinatorial property of supermodularity.
We also show that every hard three-valued {\sc Max CSP} problem
contains, in a certain specified sense, one of the two basic hard
{\sc Max CSP} problems which are the {\sc Maximum $k$-colourable
subgraph} problems for $k=2,3$.
\end{abstract}

\medskip


\bigskip

\section{Introduction and Related Work}

Many combinatorial optimization problems are $\NP$-hard, and the use of
approximation algorithms is one of the most prolific techniques to
deal with $\NP$-hardness. However, hard optimization problems
exhibit different behaviour with respect to approximability, and
complexity theory for approximation is now a well-developed
area~\cite{Ausiello99:complexity}.

Constraint satisfaction problems (CSPs) have always played a
central role in this direction of research, since the CSP
framework contains many natural computational problems, for
example, from graph theory and propositional logic. Moreover,
certain CSPs were used to build foundations for the theory of
complexity for optimization
problems~\cite{Papadimitriou91:optimization}, and some CSPs
provided material for the first optimal inapproximability
results~\cite{Hastad01:optimal} (see also
survey~\cite{Trevisan04:inapproximability}). In a CSP, informally
speaking, one is given a finite collection of constraints on
overlapping sets of variables, and the goal is to decide whether
there is an assignment of values from a given domain to the
variables satisfying all constraints (decision problem) or to find
an assignment satisfying maximum number of constraints
(optimization problem). In this paper we will focus on the
optimization problems, which are known as {\em maximum constraint
satisfaction} problems, {\sc Max CSP} for short. The most
well-known examples of such problems are {\sc Max $k$-Sat} and
{\sc Max Cut}. Let us now formally define these problems.

Let $D$ denote a {\em finite} set with $|D|>1$. Let $R^{(m)}_D$
denote the set of all $m$-ary predicates over $D$, that is,
functions from $D^m$ to $\{0,1\}$, and let
$R_D=\bigcup_{m=1}^{\infty} R^{(m)}_D$. Also, let $\Z$ denote the
set of all non-negative integers.

\begin{definition}
A {\em constraint} over a set of variables $V=\{x_1,x_2,\ldots,x_n\}$
is an expression of the form $f({\bf x})$ where

\begin{itemize}
\item
$f \in R^{(m)}_D$ is called the {\em constraint predicate}; and

\item
${\bf x} = (x_{i_1},\ldots,x_{i_m})$ is called the {\em constraint scope}.
\end{itemize}

The constraint $f$ is said to be {\em satisfied} on a tuple
${\bf a}=(a_{i_1},\ldots,a_{i_m}) \in D^m$ if $f({\bf a})=1$.
\end{definition}

\begin{definition}
For a finite $\F\sse R_D$, an instance of $\MCSP\F$ is a
pair $(V,C)$ where
\begin{itemize}
\item $V=\{x_1,\ldots,x_n\}$ is a set of variables taking their
values from the set $D$;

\item
$C$ is a collection of constraints
$f_1({\bf x}_1),\ldots,f_q({\bf x}_q)$
over $V$, where $f_i \in \F$ for all $1 \leq i \leq q$.
\end{itemize}
The goal is to find an assignment $\phi:V\rightarrow D$ that
maximizes the number of satisfied constraints,
that is, to maximize the function $f:D^n \rightarrow \Z$, defined by
$f(x_1,\ldots,x_n)=\sum_{i=1}^q f_i({\bf x}_i)$.
If the constraints have (positive integral) weights $\varrho_i$,
$1\le i\le q$,
then the goal is to maximize the total weight of satisfied constraints,
to maximize the function $f:D^n \rightarrow \Z$, defined by
$f(x_1,\ldots,x_n)=\sum_{i=1}^q \varrho_i\cdot f_i({\bf x}_i)$
\end{definition}

Note that throughout the paper the values 0 and 1 taken by any
predicate will be considered, rather unusually, as integers, not
as Boolean values, and addition will always denote the addition of
integers. It easy to check that, in the Boolean case, our problem
coincides with the {\sc Max CSP} problem considered
in~\cite{Creignou95:maximum,Creignouetal:siam01,Khanna01:approximability}.
We say that a predicate is non-trivial if it is not identically 0.
Throughout the paper, we assume that $\F$ is finite and contains
only non-trivial predicates.

{\em Boolean} constraint satisfaction problems (that is, when
$D=\{0,1\}$) are by far better studied~\cite{Creignouetal:siam01}
than the non-Boolean version. The main reason is, in our opinion,
that Boolean constraints can be conveniently described by
propositional formulas which provide a flexible and easily
manageable tool, and which have been extensively used in
complexity theory from its very birth. Moreover, Boolean CSPs
suffice to represent a number of well-known problems and to obtain
results clarifying the structure of complexity for large classes
of interesting problems~\cite{Creignouetal:siam01}. In particular,
Boolean CSPs were used to provide evidence for one of the most
interesting phenomena in complexity theory, namely that
interesting problems belong to a small number of complexity
classes~\cite{Creignouetal:siam01}, which cannot be taken for
granted due to Ladner's theorem. After the celebrated work of
Schaefer~\cite{Schaefer78:complexity} presenting a tractable
versus $\NP$-complete dichotomy for Boolean decision CSPs, many
classification results have been obtained (see,
e.g.,~\cite{Creignouetal:siam01}), most of which are dichotomies.
In particular, a dichotomy in complexity and approximability for
Boolean {\sc Max CSP} has been obtained by
Creignou~\cite{Creignou95:maximum}, and it was slightly refined
in~\cite{Khanna01:approximability} (see
also~\cite{Creignouetal:siam01}).

Many papers on various versions of Boolean CSPs mention studying
non-Boolean CSPs as a possible direction of future research, and
additional motivation for it, with an extensive discussion, was
given by Feder and Vardi~\cite{Feder98:monotone}. Non-Boolean CSPs
provide a much wider variety of computational problems. Moreover,
research in non-Boolean CSPs leads to new sophisticated algorithms
(e.g.,~\cite{Bulatov02:Mal'tsev}) or to new applications of known
algorithms (e.g.,~\cite{Cohen04:maxcsp}). Dichotomy results on
non-Boolean CSPs give a better understanding of what makes a
computational problem tractable or hard, and they give a more
clear picture of the structure of complexity of problems, since
many facts observed
in Boolean CSPs appear to be special cases of more general
phenomena. Notably, many appropriate tools for studying non-Boolean
CSPs have not been discovered until recently. For example,
universal algebra tools have proved to be very fruitful when
working with decision and counting
problems~\cite{Bulatov02:dichotomy,Bulatov03:conservative,Bulatov03:counting,Cohen:etal:jacm1997}
while ideas from combinatorial optimization and operational
research have been recently suggested for optimization
problems~\cite{Cohen04:maxcsp}.

The {\sc Max-CSP} framework has been well-studied in the Boolean
case. Many fundamental results have been obtained, concerning both
complexity classifications and approximation properties (see,
e.g.,~\cite{Creignou95:maximum,Creignouetal:siam01,Hastad01:optimal,Jonsson00:boolean,Khanna01:approximability,Zwick98:3sat}).
In the non-Boolean case, a number of results have been obtained
that concern exact (superpolynomial) algorithms or approximation
properties (see,
e.g.,~\cite{Datar03:combinatorial,Engebretsen04:non-approx,Engebretsen04:constraint,Serna98:approximability}).
The main research problem we will look at in this paper is the
following.

\begin{problem}\label{problem}
Classify the problems $\MCSP\F$ with respect to approximability.
\end{problem}

It is known that, for any $\F$, $\MCSP\F$ is an $\NPO$ problem
that belongs to the complexity class $\APX$. In other words, for any
$\F$, there is a polynomial-time approximation algorithm for
$\MCSP\F$ whose performance is bounded by a constant.

For the Boolean case, Problem~\ref{problem} was solved
in~\cite{Creignou95:maximum,Creignouetal:siam01,Khanna01:approximability}.
It appears that a Boolean $\MCSP\F$ also exhibits a dichotomy in
that it either is solvable exactly in polynomial time or else does
not admit a PTAS (polynomial-time approximation scheme) unless
\P=\NP. These papers also describe the boundary between the two
cases.

In this paper we solve the above problem for the case $|D|=3$ by
showing that $\MCSP\F$ is solvable exactly in polynomial time if,
after removing redundant values, if there are any, from the domain
(that is, taking the core), all predicates in $\F$ are
supermodular with respect to some linear ordering of the reduced
domain (see definitions in Section~\ref{supmoddefsection}) or else
the problem is $\APX$-complete. Experience shows that non-Boolean
constraint problems are much more difficult to classify, and hence
we believe that the techniques used in this paper can be further
extended to all finite domains $D$. A small technical difference
between our result and that of~\cite{Khanna01:approximability} is
that we allow repetitions of variables in constraints, as
in~\cite{Creignouetal:siam01}. Similarly
to~\cite{Creignouetal:siam01,Khanna01:approximability}, weights do
not play much role, since the tractability part of our result
holds for the weighted case, while the hardness part is true in
the unweighted case even if repetitions of constraints in
instances are disallowed. Our result uses a combinatorial property
of supermodularity which is a well-known source of tractable
optimization
problems~\cite{Burkard96:monge,Fujishige91:submodular,Topkis98:book},
and the technique of strict
implementations~\cite{Creignouetal:siam01,Khanna01:approximability}
which allows one to show that an infinite family of problems can
express, in a regular way, one of a few basic hard problems. We
remark that the idea to use supermodularity in the analysis of the
complexity of $\MCSP\F$ is very new, and has not been even
suggested in the literature prior to~\cite{Cohen04:maxcsp}.
Generally, it has been known for a while that the property of
supermodularity allows one to solve many maximization problems in
polynomial
time~\cite{Burkard96:monge,Fujishige91:submodular,Topkis98:book};
however, our result is surprising in that supermodularity appears
to be the {\em only} source of tractability for $\MCSP\F$. In the
area of approximability, examples of other works, where hardness
results are obtained for large families of problems
simultaneously,
include~\cite{Lund93:subgraph,Zuckerman96:versions}.

The only other known complete dichotomy result on a non-Boolean
constraint problem (that is, with {\em no} restrictions on $\F$)
is the theorem of Bulatov~\cite{Bulatov02:dichotomy}, where the
complexity of the standard decision problem CSP on a three-element
domain is classified. Despite the clear similarity in the settings
and also in the outcomes (full dichotomy in both cases), we note
that none of the universal-algebraic techniques used
in~\cite{Bulatov02:dichotomy} can possibly be applied in the study
of {\sc Max CSP} because the main algebraic constructions which
preserve the complexity of decision problems can be easily shown
not to do this in the case of optimization problems. Another
similarity between Bulatov's result and our theorem is that the
proof is broken down to a (relatively) large number of cases. We
believe that this is caused either by insufficiently general
methods or, more likely, by significant variation in structure of
the problems under consideration, where a large number of cases is
probably an unavoidable feature of complete classifications.

The structure of the paper is as follows: Section~\ref{prelimsec}
contains definitions of approximation complexity classes and
reductions, descriptions of our reduction techniques, and the
basics of supermodularity. Section~\ref{mainsec} contains the
proof of the main theorem of the paper.
Finally, Section~\ref{conclsec} contains a discussion of the work
we have done and of possible future work.

\section{Preliminaries}\label{prelimsec}

This section is subdivided into two parts. The first one contains
basic definitions on complexity of approximation and our reduction
techniques, while the second one introduces the notion of
supermodularity and discusses the relevance of this notion in the
study of {\sc Max CSP}.

\subsection{Approximability}

\subsubsection{Definitions}

A {\em combinatorial optimization problem} is defined over a set
of {\em instances} (admissible input data); each instance $\I$ has
a finite set ${\sf sol}(\I)$ of {\em feasible solutions}
associated with it. The {\em objective function} is, given an
instance $\I$, to find a feasible solution of {\em optimum} value.
The optimal value is the largest one for {\em maximization}
problems and the smallest one for {\em minimization} problems. A
combinatorial optimization problem is said to be an $\NP$
optimization ($\NPO$) problem if instances and solutions can be
recognized in polynomial time, solutions are polynomial-bounded in
the input size, and the objective function can be computed in
polynomial time (see, e.g.,~\cite{Ausiello99:complexity}).

\begin{definition}[performance ratio]
A solution $s$ to an instance $\mathcal{I}$ of an {\bf NPO}
problem $\Pi$ is $r$-approximate if it has value $Val$ satisfying
$$
\max{\{ \frac{Val}{Opt(\I)},\frac{Opt(\I)}{Val} \} }\le r,
$$
where $Opt(\I)$ is the optimal value for a solution to $\I$. An
approximation algorithm for an $\NPO$ problem $\Pi$ has {\em
performance ratio} $\mathcal{R}(n)$ if, given any instance $\I$ of
$\Pi$ with $|\I|=n$, it outputs an $\mathcal{R}(n)$-approximate
solution.
\end{definition}

\begin{definition}[complexity classes]
{\bf PO} is the class of {\bf NPO} problems that can be solved (to
optimality) in polynomial time. An {\bf NPO} problem $\Pi$ is in
the class $\APX$ if there is a polynomial time approximation
algorithm for $\Pi$ whose performance ratio is bounded by a
constant.
\end{definition}

Completeness in $\APX$ is defined using an appropriate reduction,
called $AP$-reduction. Our definition of this reduction
follows~\cite{Creignouetal:siam01,Khanna01:approximability}.

\begin{definition}[$AP$-reduction, $\APX$-completeness]
An $\NPO$ problem $\Pi_1$ is said to be {\em $AP$-reducible} to an
$\NPO$ problem $\Pi_2$ if two polynomial-time computable functions
$F$ and $G$ and a constant $\alpha$ exist such that
\begin{enumerate}
\item for any instance $\I$ of $\Pi_1$, $F(\I)$ is an instance of
$\Pi_2$;

\item for any instance $\I$ of $\Pi_1$, and any feasible solution
$s'$ of $F(\I)$, $G(\I,s')$ is a feasible solution of $\I$;

\item for any instance $\I$ of $\Pi_1$, and any $r\ge 1$, if $s'$
is an $r$-approximate solution of $F(\I)$ then $G(\I,s')$ is an
$(1+(r-1)\alpha+o(1))$-approximate solution of $\I$ where the
$o$-notation is with respect to $|\I|$.
\end{enumerate}

An $\NPO$ problem $\Pi$ is {\em $\APX$-hard} if every problem in
$\APX$ is $AP$-reducible to it. If, in addition, $\Pi$ is in
$\APX$ then $\Pi$ is called {\em $\APX$-complete}.
\end{definition}

It is a well-known fact (see, e.g., Section
8.2.1~\cite{Ausiello99:complexity}) that $AP$-reductions compose.
It is known that {\sc Max CSP}$({\mathcal F})$ belongs to {\bf
APX} for every ${\mathcal F}$~\cite{Cohen04:maxcsp}, and a
complete classification of the complexity of {\sc Max
CSP}$({\mathcal F})$ for a two-element set $D$ was obtained
in~\cite{Khanna01:approximability}; we will give it in
Subsection~\ref{supmoddefsection}. We shall now give an example of
an $\APX$-complete problem which will be used extensively in this
paper.

\begin{exmp}\label{maxkcol}
Given a graph $G=(V,E)$,
the {\sc Maximum $k$-colourable Subgraph} problem, $k\ge 2$,
is the problem of maximizing $|E'|$, $E'\sse E$,
such that the graph $G'=(V,E')$ is $k$-colourable.
This problem is known to be \APX-complete problem
(it is Problem GT33 in~\cite{Ausiello99:complexity}).
Let $neq_k$ denote the binary disequality predicate on $\{0,1\zd k-1\}$,
$k\ge 2$, that is, $neq_k(x,y)=1 \Leftrightarrow x\ne y$.
The problem $\MCSP{\{neq_k\}}$ is slightly more general than the
{\sc Maximum $k$-colourable Subgraph} problem. To see this, think of
vertices of a given graph as of variables, and apply the predicate
to every pair of variables $x,y$ such that $(x,y)$ is an edge in the graph.

If we allow weights on edges in graphs and on constraints then the
problems are precisely the same. For unweighted problems, the
$\MCSP{\{neq_k\}}$ is slightly more general because one can have
constraints $neq_k(x,y)$ and $neq_k(y,x)$ in the same instance. In
any case, it follows that the problem $\MCSP{\{neq_k\}}$ is
\APX-complete.
\end{exmp}

Interestingly, the problems $\MCSP{\{neq_k\}}$, $k=2,3$,
will be the only basic hard problems for the case $|D|\le 3$.
We will show that,
for all other \APX-complete problems $\MCSP\F$, the set
$\F$ can express, in a certain regular approximability-preserving way,
one of the predicates $neq_2$, $neq_3$.

\subsubsection{Reduction techniques}

The basic reduction technique in our $\APX$-completeness proofs is
based on {\em strict implementations},
see~\cite{Creignouetal:siam01,Khanna01:approximability} where this
notion was defined and used only for the Boolean case. We will
give this definition in a different form from that
of~\cite{Creignouetal:siam01,Khanna01:approximability}, but it can
easily be checked to be equivalent to the original one (in the
case $|D|=2$).

\begin{definition}\label{strictdef}
Let $Y=\{y_1,\ldots,y_m\}$ and $Z=\{z_1,\ldots,z_n\}$ be two
disjoint sets of variables. The variables in $Y$ are called
primary and the variables in $Z$ auxiliary. The set $Z$ may be
empty. Let $g_1({\bf y}_1),\ldots,g_s({\bf y}_s)$, $s > 0$, be
constraints over $Y \cup Z$. If $g(y_1,\ldots,y_m)$ is a predicate
such that the equality
$$
g(y_1,\ldots,y_m) + (\alpha -1) = \max_{Z}{\sum_{i=1}^{s}{g_i({\bf y}_i)}}
$$
is satisfied for all $y_1,\ldots,y_m$, and some fixed $\alpha \in
\Z$, then this equality is said to be a {\em strict
\mbox{$\alpha$-implementation}} of $g$ from $g_1\zd g_s$.

\end{definition}
We use $\alpha-1$ rather than $\alpha$ in the above equality to
ensure that this notion coincides with the original notion of a
strict $\alpha$-implementation for Boolean
constraints~\cite{Creignouetal:siam01,Khanna01:approximability}.

We say that a collection of predicates ${\cal F}$ {\em strictly
implements} a predicate $g$ if,  for some $\alpha\in\Z$, there
exists a strict $\alpha$-implementation of $g$ using predicates
only from ${\cal F}$.
In this case we write $\F
\stackrel{s\phantom{8pt}}{\Longrightarrow_\alpha} f$. It is not
difficult to show that if $f$ can be obtained from $\F$ by a
series of strict implementations then it can also be obtained by a
single strict implementation. In this paper, we will use about 60
(relatively) short strict implementations for the case when
$|D|=3$. Each of them can be straightforwardly verified by hand,
or (better still) by a simple computer program\footnote{An example
of such a program can be obtained from the authors or be anonymously
downloaded from
\texttt{http://www.ida.liu.se/\~{}mikkl/verifier/}.}.

\begin{lemma}\label{strict}
If $\F$ strictly implements a predicate $f$, and
$\MCSP{\F\cup\{f\}}$ is $\APX$-complete, then
$\MCSP\F$ is \APX-complete as well.
\end{lemma}

\begin{proof}
We need to show that $\MCSP{\F\cup\{f\}}$ is $AP$-reducible to
$\MCSP\F$. For the case $|D|=2$, this was proved in Lemma
5.18 of \cite{Creignouetal:siam01}. To show this for the general
case, repeat the proof of the above mentioned lemma
from~\cite{Creignouetal:siam01}, replacing 2 by $|D|$.
\end{proof}

Lemma~\ref{strict} will be used as follows in our
\APX-completeness proofs: if $\F'$ is a fixed finite collection of
predicates each of which can be strictly implemented by $\F$ then
we can assume that $\F'\sse \F$. For example, if $\F$ contains a
binary predicate $f$ then we can assume, at any time when it is
convenient, that $\F$  also contains $f'(x,y)=f(y,x)$, since this
equality is a strict 1-implementation of $f'$.

\begin{exmp}\label{maxcutexample}
The {\sc (Simple) Max Cut} problem is the problem of partitioning
the set of vertices of a given undirected graph into two subsets
so as to maximize the number of edges with ends being in different
subsets. This problem is the same as {\sc Maximum 2-colourable
Subgraph} (see Example~\ref{maxkcol}), and hence it is
$\APX$-complete (see Problem ND14
in~\cite{Ausiello99:complexity}). As was mentioned in
Example~\ref{maxkcol}, this problem is essentially the same as
$\MCSP{\{neq_2\}}$. Let $f_{dicut}$ be the binary predicate on $\{0,1\}$
such that $f_{dicut}(x,y)=1 \Leftrightarrow x=0,y=1$. Then $\MCSP{\{f_{dicut}\}}$
is essentially the problem {\sc Max Dicut} (see problem ND16
in~\cite{Ausiello99:complexity}), which is the problem of
partitioning the vertices of a digraph into two subsets $V_0$ and
$V_1$ so as to maximize the number of arcs going from $V_0$ to
$V_1$. This problem is known to be $\APX$-complete as well, and
this can be proved by exhibiting a strict 1-implementation from
$f_{dicut}$ to $neq_2$. Here it is: $neq_2(x,y)=f_{dicut}(x,y)+f_{dicut}(y,x)$.
\end{exmp}

For a subset $D'\sse D$, let $u_{D'}$ denote the predicate such
that $u_{D'}(x)=1$ if and only if $x \in D'$. Let ${\cal
U}_D=\{u_{D'} \; | \; \emptyset\ne D' \subseteq D\}$, that is,
$\U_D$ is the set of all non-trivial unary predicates on $D$. We
will now give two more examples of strict implementations that
will be used later in our proofs.

\begin{exmp}\label{3to1example}
Let $D=\{0,1,2\}$, and $g_i$, $i=0,1,2$, be the binary predicates
on $D$ defined by the following rule: $g_i(x,y)=1 \Leftrightarrow
(x=y=i \mbox{ or } x,y\in D\setminus\{i\})$. We will show that
$\F=\{g_0,g_1,g_2\}\cup \U_D$ strictly implements the binary
predicate $g(x,y)$ such that $g(x,y)=1 \Leftrightarrow x=0,y=1$.
Indeed, one can check that the following is a strict
5-implementation:
$$
g(x,y)+4=\max_{z,w}[g_{0}(x,z)+g_1(y,w)+g_2(z,w)+u_{\{0\}}(z)+u_{\{1,2\}}(w)].$$
\end{exmp}

\begin{exmp}\label{neqfromeq}
In this example, we will show that the predicate $neq_3$ can be
strictly implemented from the binary equality predicate $eq_3$ and
all unary predicates on $D=\{0,1,2\}$. We will use three
additional binary predicates $f_1,f_2,f_3$ defined as follows:
\begin{enumerate}
\item[] \mbox{$f_1(x,y)=1 \Leftrightarrow x\le y$},

\item[] \mbox{$f_2(x,y)=1 \Leftrightarrow (x,y)=(1,2)$},

\item[] \mbox{$f_3(x,y)=1 \Leftrightarrow (x,y)\in
\{(1,0),(1,2),(2,0)\}$}.
\end{enumerate}

Then it can be checked that the following equalities hold:

\begin{eqnarray*}
f_1(x,y)+3 & = & \max_{z,w}[eq_3(z,w)+eq_3(z,y)+eq_3(w,x)+u_{\{2\}}(z)+\\
& &
+u_{\{1\}}(w)+u_{\{0\}}(x)];\\ \\
f_2(x,y)+5 & = & \max_{z,w}[f_1(z,w)+f_1(w,y)+f_1(x,z)+u_{\{0,1\}}(z)+\\
& &
+u_{\{0,2\}}(w)+u_{\{1,2\}}(x)];\\ \\
f_3(x,y)+2 & = &
\max_{z,w}[f_2(z,w)+f_2(z,x)+f_2(w,z)+f_2(w,y)+f_2(x,w)+ \\ \\
& &
+f_2(x,y)+f_2(y,z)+
  +u_{\{0\}}(y)];\\ \\
neq_3(x,y) & = & f_3(x,y)+f_3(y,x).
\end{eqnarray*}
%
As mentioned above, a chain of strict implementations can be
replaced by a single strict implementation. Since
$\MCSP{\{neq_3\}}$ is $\APX$-complete by Example~\ref{maxkcol},
Lemma~\ref{strict} imply that the problem
$\MCSP{\{eq_3\}\cup \U_D}$ is $\APX$-complete as well. Note that
this result was first proved in~\cite{Cohen04:maxcsp}.
\end{exmp}

\medskip

Another notion which we will use in our hardness proofs is
the notion of a core for a set of predicates.
In the case when  $\F$ consists of a single binary predicate $h$,
this notion coincides with the usual notion of a core of the directed
graph whose arcs are specified by $h$.

\begin{definition}
An {\em endomorphism} of $\F$ is a unary operation $\pi$ on $D$ such that,
for all $f\in \F$  and all $(a_1\zd a_m) \in D^m$, we have
$f(a_1\zd a_m)=1 \Rightarrow f(\pi(a_1)\zd \pi(a_m))=1$.
We will say that $\F$ is a {\em core} if every endomorphism of $\F$
is injective (i.e., a permutation).

If $\pi$ is an endomorphism of $\F$ with a minimal image $im(\pi)=D'$
then a core of $\F$, denoted $core(\F)$,
is the subset $\{f|_{D'}\mid f\in \F\}$ of $R_{D'}$.
\end{definition}
The intuition here is that if $\F$ is not a core then
it has a non-injective endomorphism $\pi$, which implies that, for
every assignment $\phi$, there is another assignment $\pi\phi$
that satisfies all constraints satisfied by $\phi$ and uses only a
restricted set of values, so the problem is equivalent to a
problem over this smaller set. As in the case of graphs, all cores
of $\F$ are isomorphic, so one can speak about {\em the} core of
$\F$. The following rather simple lemma will be frequently used in
our proofs.

\begin{lemma}\label{core}
If $\F'=core(\F)$ and $\MCSP{\F'}$ is \APX-complete then so is $\MCSP\F$.
\end{lemma}

\begin{proof}
We produce an $AP$-reduction from $\MCSP{\F'}$ to $\MCSP{\F}$. We
may assume that the endomorphism $\pi:D\rightarrow D'$ is the
identity on $D'$, since if it is not, then one of its powers is
such an endomorphism. We will now describe functions $F$ and $G$
necessary for the reduction. The function $F$ takes an instance of
$\MCSP{\F'}$ and replaces every predicate $f|_{D'}$ in it by $f$.
If $\I$ is an instance of $\MCSP{\F'}$, with the set $V$ of
variables, and $s'$ is a feasible solution of $F(\I)$ (that is, an
assignment $V\rightarrow D$) then $G(F(\I),s')=s$ defined by
$s(x)=\pi(s'(x))$ for all $x\in V$. It is easy to see that $s$ is
also a feasible solution for $\I$. Finally, note that, since $\pi$
is an endomorphism, $s$ satisfies every constraint satisfied by
$s'$; in particular, we have $Opt(\I)=Opt(F(\I))$. Hence, if $s'$
is an $r$-approximate solution for $F(\I)$ then $s$ is an
$r$-approximate solution for $\I$, so we can choose $\alpha=1$ in
the definition of $AP$-reducibility.
\end{proof}

\begin{exmp}
Let $f$ be the binary predicate on $\{0,1\}$ considered in
Example~\ref{maxcutexample}, and  $g$ the binary predicate on
$\{0,1,2\}$ considered in Example~\ref{3to1example}. It is easy to
see that $\{f\}$ is the core of $\{g\}$ where the corresponding
endomorphism is given by $\pi(0)=0, \pi(1)=\pi(2)=1$. Since
$\MCSP{\{f\}}$ is $\APX$-complete, Lemma~\ref{core} implies that
$\MCSP{\{g\}}$ is $\APX$-complete as well. Now note that this also
proves that $\MCSP{\{g_0,g_1,g_2\}\cup
\U_{\{0,1,2\}}}$, as considered in Example~\ref{3to1example}, is
$\APX$-complete.
\end{exmp}

\subsection{Supermodularity}
\label{supmoddefsection}

In this section we discuss the well-known combinatorial algebraic
property of supermodularity~\cite{Topkis98:book} which will play a
crucial role in classifying the approximability of {\sc Max CSP}
problems.

A partial order on a set $D$ is called a
{\em lattice order} if, for every $x,y \in D$, there exists
a greatest lower bound $x \sqcap y$ and a least upper bound
$x \sqcup y$. The corresponding
algebra ${\cal L}=(D,\sqcap,\sqcup)$ is called a {\em lattice}.
For tuples ${\bf a}=(a_1,\ldots,a_n)$, ${\bf b}=(b_1,\ldots,b_n)$
in $D^n$, let ${\bf a} \sqcap {\bf b}$ and
${\bf a} \sqcup {\bf b}$ denote the tuples
$(a_1 \sqcap b_1,\ldots,a_n \sqcap b_n)$ and
$(a_1 \sqcup b_1,\ldots,a_n \sqcup b_n)$, respectively.

\begin{definition}
Let ${\cal L}$ be a lattice on $D$. A function
$f:D^n \rightarrow \Z$ is called {\em supermodular}
on ${\cal L}$ if

\[f({\bf a})+f({\bf b}) \leq
f({\bf a} \sqcap {\bf b})+f({\bf a} \sqcup {\bf b}) \; \;
\mbox{for all ${\bf a},{\bf b} \in D^n$},\]
and $f$ is called {\em submodular} on ${\cal L}$
if the inverse inequality holds.
\end{definition}

We say that $\F\sse R_D$ is supermodular on $\Lat$ if every
$f\in\F$ has this property.

A finite lattice $\Lat=(D,\sqcap,\sqcup)$ is {\em distributive} if
and only if it can be represented by subsets of a set $A$, where
the operations $\sqcap$ and $\sqcup$ are interpreted as
set-theoretic intersection and union, respectively. Totally
ordered lattices, or {\em chains}, will be of special interest in
this paper. Note that, for chains, the operations $\sqcap$ and
$\sqcup$ are simply $\min$ and $\max$. Hence, the supermodularity
property for an $n$-ary predicate $f$ on a chain is expressed as
follows:
\[f(a_1\zd a_n)+f(b_1\zd b_n) \leq \]
\[f(\min(a_1,b_1)\zd \min(a_n,b_n))+f(\max(a_1,b_1)\zd
\max(a_1,b_1))\]
for all $a_1\zd a_n,b_1\zd b_n$.

\begin{exmp}\label{supmodexample}$~$

1) The binary equality predicate $eq_3$ is not supermodular on any
chain on $\{0,1,2\}$. Take, without loss of generality, the chain
$0<1<2$. Then $$eq_3(1,1)+eq_3(0,2)=1\not\le 0=
eq_3(0,1)+eq_3(1,2).$$

2) Reconsider the predicates $neq_2$ and $f_{dicut}$ from
Example~\ref{maxcutexample}. It is easy to check that neither of
them is supermodular on any chain on $\{0,1\}$.

3) Fix a chain on $D$ and let ${\bf a}, {\bf b}$ be arbitrary
elements of $D^2$. Consider the binary predicate $f_{{\bf a}}$,
$f^{{\bf b}}$ and $f_{{\bf a}}^{{\bf b}}$ defined by the rules
\begin{eqnarray*}
f_{{\bf a}}(x,y)=1 & \Leftrightarrow & (x,y)\le {\bf a},\\
f^{{\bf b}}(x,y)=1 & \Leftrightarrow & (x,y)\ge {\bf b},\\
f_{{\bf a}}^{{\bf b}}(x,y)=1 & \Leftrightarrow & (x,y)\le {\bf a}
\mbox{ or } (x,y)\ge {\bf b},
\end{eqnarray*}
where the order on $D^2$ is component-wise.  It is easy to check
that every predicate of one of the forms above is
supermodular on the chain. Note that such predicates were
considered in~\cite{Cohen04:maxcsp} where they were called
generalized 2-monotone. We will see later in this subsection that
such predicates are generic supermodular binary predicates on a
chain.
\end{exmp}

We will now make some very simple, but useful, observations.

\begin{observation}\label{spmodobs}
$~$

\begin{enumerate}
\item Any chain is a distributive lattice.

\item Any lattice on a three-element set is a chain.

\item Any unary predicate on $D$ is supermodular on any chain on
$D$.

\item  A predicate is supermodular on a chain if and only if it
is supermodular on its dual chain (obtained by reversing the
order).
\end{enumerate}
\end{observation}

The tractability part of our classification is contained in the
following result:

\begin{theorem}[\cite{Cohen04:maxcsp}]\label{tract}
If $\F$ is supermodular on some distributive lattice on $D$, then
weighted $\MCSP\F$ is in $\PO$.
\end{theorem}

Given a binary predicate $f:D^2 \rightarrow \{0,1\}$, we will
often use a $|D| \times |D|$ 0/1-matrix $M$ to represent $f$:
$f(x,y)=1$ if and only if $M_{xy}=1$. Note that this matrix is
essentially the table of values of the predicate. For example,
some binary predicates on $D=\{0,1,2\}$ that are supermodular on
the chain $0<1<2$ are listed in Fig.~\ref{spmodlist}. Matrices for all
other binary predicates that are supermodular on $0<1<2$ can be
obtained from those in the list or from the trivial binary
predicate by transposing matrices (which corresponds to swapping
arguments in a predicate) and by replacing some all-0 rows by
all-1 rows, and the same for all-0 columns (but not for both rows
and columns at the same time). This can be shown by using Lemma
2.3 of \cite{Burkard96:monge} or by direct exhaustive
(computer-assisted) search.
\begin{figure}[t]
\setlength{\unitlength}{5.42mm}
\begin{center}
\begin{picture}(2,3)(0,0)
\put(0.2,0.85){\footnotesize $h_{1}$}
\put(1,1.3){\footnotesize 000}
\put(1,0.8){\footnotesize 000}
\put(1,0.3){\footnotesize 001}
\end{picture}
\begin{picture}(2,3)(0,0)
\put(0.2,0.85){\footnotesize $h_{2}$}
\put(1,1.3){\footnotesize 000}
\put(1,0.8){\footnotesize 000}
\put(1,0.3){\footnotesize 011}
\end{picture}
\begin{picture}(2,3)(0,0)
\put(0.2,0.85){\footnotesize $h_{3}$}
\put(1,1.3){\footnotesize 000}
\put(1,0.8){\footnotesize 011}
\put(1,0.3){\footnotesize 011}
\end{picture}
\begin{picture}(2,3)(0,0)
\put(0.2,0.85){\footnotesize $h_{4}$}
\put(1,1.3){\footnotesize 100}
\put(1,0.8){\footnotesize 000}
\put(1,0.3){\footnotesize 000}
\end{picture}
\begin{picture}(2,3)(0,0)
\put(0.2,0.85){\footnotesize $h_{5}$}
\put(1,1.3){\footnotesize 100}
\put(1,0.8){\footnotesize 000}
\put(1,0.3){\footnotesize 001}
\end{picture}
\begin{picture}(2,3)(0,0)
\put(0.2,0.85){\footnotesize $h_{6}$}
\put(1,1.3){\footnotesize 100}
\put(1,0.8){\footnotesize 000}
\put(1,0.3){\footnotesize 011}
\end{picture}
\begin{picture}(2,3)(0,0)
\put(0.2,0.85){\footnotesize $h_{7}$}
\put(1,1.3){\footnotesize 100}
\put(1,0.8){\footnotesize 011}
\put(1,0.3){\footnotesize 011}
\end{picture}
\begin{picture}(2,3)(0,0)
\put(0.2,0.85){\footnotesize $h_{8}$}
\put(1,1.3){\footnotesize 100}
\put(1,0.8){\footnotesize 100}
\put(1,0.3){\footnotesize 000}
\end{picture}
\begin{picture}(2,3)(0,0)
\put(0.2,0.85){\footnotesize $h_{9}$}
\put(1,1.3){\footnotesize 100}
\put(1,0.8){\footnotesize 100}
\put(1,0.3){\footnotesize 001}
\end{picture}
\begin{picture}(2,3)(0,0)
\put(0.1,0.85){\footnotesize $h_{10}$}
\put(1,1.3){\footnotesize 100}
\put(1,0.8){\footnotesize 100}
\put(1,0.3){\footnotesize 011}
\end{picture}
\begin{picture}(2,3)(0,0)
\put(0.1,0.85){\footnotesize $h_{11}$}
\put(1,1.3){\footnotesize 100}
\put(1,0.8){\footnotesize 101}
\put(1,0.3){\footnotesize 001}
\end{picture}
\begin{picture}(2,3)(0,0)
\put(0.1,0.85){\footnotesize $h_{12}$}
\put(1,1.3){\footnotesize 110}
\put(1,0.8){\footnotesize 110}
\put(1,0.3){\footnotesize 000}
\end{picture}
\begin{picture}(2,3)(0,0)
\put(0.1,0.85){\footnotesize $h_{13}$}
\put(1,1.3){\footnotesize 110}
\put(1,0.8){\footnotesize 110}
\put(1,0.3){\footnotesize 001}
\end{picture}

\end{center}
\caption{A list of binary predicates on $\{0,1,2\}$
that are supermodular on the chain $0<1<2$.
The predicates are represented by matrices, the order of indices
being also $0<1<2$.}
\label{spmodlist}
\end{figure}
Note that all predicates in Fig.~\ref{spmodlist} have the form
described in Example~\ref{supmodexample}(3). For example, $h_2$ is
$f^{(2,1)}$ and $h_9$ is $f_{(1,0)}^{(2,2)}$.

The property of supermodularity can be used to classify the
approximability of Boolean problems $\MCSP\F$ (though, originally
the classification was
obtained and stated~\cite{Creignou95:maximum,Creignouetal:siam01,%
Khanna01:approximability} without using this property).
It is easy to see that $\F\sse R_{\{0,1\}}$
is not a core if and only if $f(a\zd a)=1$
for some $a\in\{0,1\}$ and all $f\in\F$,
in which case $\MCSP\F$ is trivial.

\begin{theorem}[\cite{Cohen04:maxcsp,Creignouetal:siam01}]\label{Boolean}
Let $D=\{0,1\}$ and $\F\sse R_D$ be a core. If $\F$ is supermodular
on some chain on $D$ then $\MCSP\F$ belongs to $\PO$.
Otherwise, $\MCSP\F$ is $\APX$-complete.
\end{theorem}

\begin{remark} It was shown in Lemma 5.37 of \cite{Creignouetal:siam01}
that $\F$ can strictly implement $neq_2$ whenever $\MCSP\F$ is
$\APX$-complete in the above theorem.
\end{remark}

Combining Theorem~\ref{Boolean} with Lemma~\ref{core}, we get the
following corollary which will be used often in our
\APX-completeness proofs.

\begin{corollary}\label{2core}
If $g'$ is binary predicate on $\{0,1,2\}$ and $core(\{g'\})$ is
$\{g\}$ where $g$ is non-super\-mo\-du\-lar predicate on a
two-element subset of $D$ then $\MCSP{\{g'\}}$ is \APX-complete.
\end{corollary}

Note that there are only two (up to swapping of arguments) binary
predicates $g$ on $\{0,1\}$ such that $\{g\}$ is a core: the
predicates $neq_2$ and $f_{dicut}$ from Example~\ref{maxcutexample}. As
mentioned above, these two predicates are non-supermodular, and
$f_{dicut}$ strictly 1-implements $neq_2$.

\section{Main result}\label{mainsec}

In this section we establish a generalization of Theorem~\ref{Boolean}
to the case of a three-element domain. Throughout this section, let
$D=\{0,1,2\}$. Note that if $\F\sse R_D$ is not a core then,
by Lemma~\ref{core},
the problem $\MCSP\F$ is either trivial (if $\F$ has a constant endomorphism)
or else reduces to a similar problem over a two-element domain,
in which case Theorem~\ref{Boolean} applies.

\begin{theorem}\label{main}
Let $D=\{0,1,2\}$ and $\F\sse R_D$ be a core. If $\F$ is
supermodular on some chain on $D$ then weighted $\MCSP\F$ belongs
to $\PO$. Otherwise, unweighted $\MCSP\F$ is $\APX$-complete even
if repetitions of constraints in instances are disallowed.
\end{theorem}

\begin{proof}
The tractability part of the proof follows immediately from
Theorem~\ref{tract} (see also Observation~\ref{spmodobs}(1)).
Assume for the rest of this section that $\F$ is a core and it is
not supermodular on any chain on $D$. We will show that 
one of $neq_2$, $neq_3$ can be obtained from
$\F$ by using the following two operations:
\begin{enumerate}
\item replacing $\F$ by $\F\cup \{f\}$ where $f$ is a predicate
that can be strictly implemented from $\F$;

\item taking the core of a subset of $\F$.
\end{enumerate}
By Example~\ref{maxkcol} and Lemmas~\ref{strict} and~\ref{2core},
this will establish the result.

To improve readability, we divide the rest of the proof into 3
parts: in Subsection~\ref{smallcases}, we establish
$\APX$-completeness for some small sets $\F$ consisting of at most
two binary and several unary predicates, and also for the case
when  $\F$ contains an irreflexive non-unary predicate (see
definition below). Subsection~\ref{allunarysubsec} establishes the
result when all unary predicates are available, and
Subsection~\ref{generalsec} finishes the proof.
\end{proof}

\begin{remark}
Note that it can be checked in polynomial time whether a given
$\F$ is supermodular on some chain on $D$, if the predicates in
$\F$ are given by full tables of values or only by tuples on which
predicates take value 1.
\end{remark}

\subsection{Small cases and irreflexive predicates}
\label{smallcases}

We say that an $n$-ary predicate $f$ on $D$
is {\em irreflexive} if and only if $f(d,\ldots,d)=0$ for
all $d \in D$. It is easy to check that any
irreflexive non-trivial predicate $f$ is not supermodular on any chain
on $D$. For example, if $f$ is binary and $f(a,b)=1$ for some $a\ne b$
then $f(a,b)+f(b,a)\ge 1$, but $f(min(a,b),min(b,a))+f(max(a,b),max(b,a))=0$
due to irreflexivity.

Since a predicate $f$ is supermodular on a chain $C$ if and only
if $f$ is supermodular on its dual, we can identify chains on the
three-element set $D$ with the same middle element: let $C_i$
denote an arbitrary chain on $D$ with $i$ as its middle element.
We also define the set $\Q_i$ that consists of all binary
predicates on $D$ that are supermodular on $C_i$ but on neither of
the other two chains. For example, it is easy to check using
Fig.~\ref{spmodlist} that $\Q_1$ consists of predicates
$h_2,h_5,h_6,h_8,h_9,h_{10},h_{11}$ and the predicates obtained
from them by using the following operations:
\begin{enumerate}
\item swapping the variables (this corresponds to transposing the
tables);

\item adding a unary predicate $u(x)$ or $u(y)$ in such a way that
the sum remains to be a predicate (this corresponds to replacing
all-0 rows/columns with all-1 rows/columns).
\end{enumerate}

Recall that, for a subset $D'\sse D$, $u_{D'}$ denotes the
predicate such that $u_{D'}(x)=1$ if and only if $x \in D'$, and
${\cal U}_D=\{u_{D'} \; | \; \emptyset\ne D' \subseteq D\}$, that
is, $\U_D$ is the set of all non-trivial unary predicates on $D$.

\begin{lemma} \label{all-one}
Let $g$ be a binary predicate such that, for some
$a \in D$,  $g(x,a)=1$
for all $x \in D$. Let $g'(x,y)=0$ if $y=a$ and $g'(x,y)=g(x,y)$ otherwise.
Then, the following holds:

\begin{enumerate}
\item
for any chain on $D$, $g$ and $g'$ are supermodular
(or not) on it simultaneously; and

\item
$\{g,u_{D \setminus \{a\}}\})$ strictly implements $g'$.
\end{enumerate}
\end{lemma}
\begin{proof}
The first statement is a straightforward consequence of the
definition. To see that the second statement holds, we note that
$g'(x,y)+1= g(x,y)+u_{D \setminus \{a\}}(y)$ is a strict
2-implementation of $g'(x,y)$.
\end{proof}

We say that a predicate $g$ contains an {\em all-one column} if
there exists $a \in D$ such that $g(x,a)=1$ for all $x \in D$, and
we define {\em all-one rows} analogously. Clearly, the lemma above
holds for both all-one rows and all-one columns. The lemma will be
used in our hardness proofs as follows: if $\F$ contains $g(x,y)$
and $u_{D \setminus \{a\}}(y)$ then, by Lemma~\ref{strict}, we may
also assume that $g'\in \F$.

The following lemma contains more $\APX$-completeness results
for some problems $\MCSP\F$ where $\F$ is a small set containing
at most two binary and some unary predicates.

\begin{lemma} \label{morehardcases}
Let $f,h$ be binary predicates on $D$.
The problem $\MCSP\F$ is $\APX$-complete
if one of the following holds:

\begin{enumerate}
\item $\F=\{f\}$ and $f$ is nontrivial and irreflexive; \item
$\F=\{f\} \cup {\mathcal U}_D$ and $f$ is not supermodular on any
chain on $D$; \item $\F=\{f,h_7\} \cup {\mathcal U}_D$ where $f
\in \Q_0$ and $h_7$ is given in Fig.~\ref{spmodlist}; \item
$\F=\{f,h\} \cup {\mathcal U}_D$ and $f \in \Q_1$ and $h \in
\Q_0$; \item $\F=\{f,u_{\{0,1\}}\}$ where $f$ is such that $f(0,0)
= f(1,1) = 0$ and $f(2,2)=f(0,1) = 1$.
\end{enumerate}
\end{lemma}
\begin{proof}
The lemma is proved by providing computer-generated strict
implementations, from $\F$, of the predicate $neq_3$ (see
Example~\ref{maxkcol}) or of a binary predicate whose core is a
non-supermodular predicate on a two-element subset of $D$ (see
Corollary~\ref{2core}). In total, we give 54 implementations.

We prove only case 1 here; the other cases are similar and can be
found in the Appendix.
First, we make the list of predicates we need to consider.
There are 63 irreflexive non-trivial predicates on $D$.
We may skip all predicates whose core is a non-supermodular predicate
on a two-element subset of $D$, since we already have the result for them
(Corollary~\ref{2core}).
For every pair of predicates that can be obtained from each other by swapping
the variables (that is, $f(x,y)$ and $f'(x,y)=f(y,x)$), we can skip one of
them. By symmetry, we may skip any predicate obtained from some predicate
already in the list by renaming the elements of $D$.
Finally, we already know that the result is true for the disequality
predicate $neq_3$, so we skip that one too.
All this can be done using a computer or by hand, and the resulting list
contains only six predicates. Here are strict implementations for them.

{\flushleft
\begin{enumerate}
\item
{\normalsize $f_{1} := {\small \begin{array}{l}
011\\
001\\
000\\
\end{array}
} \stackrel{s\phantom{8pt}}{\Longrightarrow_1}$ ${\small
\begin{array}{l}
011\\
101\\
110\\
\end{array}
} = neq_3$}

{\normalsize Implementation: $neq_3=f_{1}(x,y)+f_{1}(y,x)$}

\item

{\normalsize $f_{2} := {\small \begin{array}{l}
010\\
001\\
100\\
\end{array}
} \stackrel{s\phantom{8pt}}{\Longrightarrow_1}$ ${\small
\begin{array}{l}
011\\
101\\
110\\
\end{array}
} = neq_3$}

{\normalsize Implementation: $neq_3(x,y)=f_{2}(x,y)+f_{2}(y,x)$}

\item

{\normalsize $f_{3} := {\small \begin{array}{l}
011\\
101\\
100\\
\end{array}
} \stackrel{s\phantom{8pt}}{\Longrightarrow_3}$ ${\small
\begin{array}{l}
010\\
001\\
100\\
\end{array}
} = f_{2}$}

{\normalsize Implementation:
$f_2(x,y)+2=\max_{z}[f_{3}(z,x)+f_{3}(x,y)+f_{3}(y,z)]$}

\item

{\normalsize $f_{4} := {\small \begin{array}{l}
011\\
101\\
000\\
\end{array}
} \stackrel{s\phantom{8pt}}{\Longrightarrow_3}$ ${\small
\begin{array}{l}
011\\
101\\
110\\
\end{array}
} = neq_3$}

{\normalsize Implementation:
$neq_3(x,y)+2=\max_{z}[f_{4}(z,x)+f_{4}(z,y)+f_{4}(x,y)+f_{4}(y,x)$]}

\item

{\normalsize $f_{5} := {\small \begin{array}{l}
001\\
100\\
000\\
\end{array}
} \stackrel{s\phantom{8pt}}{\Longrightarrow_3}$ ${\small
\begin{array}{l}
011\\
101\\
000\\
\end{array}
} = f_{4}$}

{\normalsize Implementation:
$f_4(x,y)+2=\max_{z,w}[f_{5}(z,w)+f_{5}(z,y)+f_{5}(w,y)+f_{5}(x,w)+f_{5}(y,z)]$}

\item

{\normalsize $f_{6} := {\small \begin{array}{l}
011\\
001\\
100\\
\end{array}
} \stackrel{s\phantom{8pt}}{\Longrightarrow_4}$ ${\small
\begin{array}{l}
011\\
101\\
100\\
\end{array}
} = f_{3}$}

{\normalsize Implementation:
$f_3(x,y)+3=\max_{z,w}[f_{6}(z,y)+f_{6}(w,z)+f_{6}(w,x)+f_{6}(x,z)+f_{6}(x,y)]$}
\end{enumerate}

}

\end{proof}

\begin{proposition} \label{irrefl}
If $h\in R_D^{(n)}$, $n\ge 2$, is nontrivial and irreflexive, then
{\sc Max CSP}$(\{h\})$ is {\bf APX}-complete.
\end{proposition}
\begin{proof}
The proof is by induction on $n$ (the arity of $h$).
The basis when $n=2$ was proved in Lemma~\ref{morehardcases}(1).
Assume that the lemma holds for $n=k$, $k \geq 2$. We show that
it holds for $n=k+1$. Assume first that
there exists $(a_1,\ldots,a_{k+1}) \in D^{k+1}$ such that
$h(a_1,\ldots,a_{k+1})=1$ and
$|\{a_1,\ldots,a_{k+1}\}| \leq k$.
We assume without loss of generality that $a_k=a_{k+1}$ and
consider the predicate $h'(x_1\zd x_k)=h(x_1,\ldots,x_k,x_k)$.
Note that this is a strict 1-implementation of $h'$, that
$h'(d,\ldots,d)=0$ for all $d \in D$, and that $h'$ is nontrivial
since $h'(a_1,\ldots,a_k)=1$. Consequently, $\MCSP{\{h'\}}$ is
{\bf APX}-complete by the induction hypothesis, and {\sc Max
CSP}$(\{h\})$ is {\bf APX}-complete, too.

Assume now that $|\{a_1,\ldots,a_{k+1}\}| = k+1$ whenever
$h(a_1,\ldots,a_{k+1})=1$. Consider the predicate $h'(x_1\zd x_k)=
\max_{y}h(x_1\zd x_k,y)$, and  note that this is a strict
1-implementation of $h'$. We see that $h'(d,\ldots,d)=0$ for all
$d \in D$ (due to the condition above) and $h'$ is non-trivial
since $h$ is non-trivial. We can once again apply the induction
hypothesis and draw the conclusion that {\sc Max CSP}$(\{h'\})$
and {\sc Max CSP}$(\{h\})$ are {\bf APX}-complete.
\end{proof}

\subsection{When all unary predicates are available} \label{allunarysubsec}

As the next step, we will prove that $\MCSP{\F\cup \U_D}$ is \APX-complete
if $\F$ is not supermodular on any chain.
As a special case of Lemma~6.3 of \cite{Burkard96:monge}, we have the
following result (see also Observation~6.1 of \cite{Burkard96:monge}).

\begin{lemma} \label{binaryenough}
An $n$-ary, $n\ge 2$, predicate $f$ is supermodular on a fixed chain $C$
if and only if the following holds: every binary predicate
obtained from $f$ by replacing any given $n-2$ variables by any constants
is supermodular on $C$.
\end{lemma}

\begin{proposition}\label{withun}
$\MCSP{\F\cup \U_D}$ is \APX-complete if $\F$ is not supermodular 
on any chain.
\end{proposition}
\begin{proof}
By our initial assumptions, $\F$ is not supermodular on any chain.
For $i=0,1,2$, let $f_i\in \F$ be {\em not} supermodular on $C_i$.
Recall that every unary predicate is supermodular on any chain.
Therefore, $f_i$ is $n$-ary where $n\ge 2$ (note that $n$ depends
on $i$). By Lemma~\ref{binaryenough}, it is possible to substitute
constants for some $n-2$ variables of $f_i$ to obtain a binary
predicate $f'_i$ which is not supermodular on $C_i$. Assume
without loss of generality that these variables are the last $n-2$
variables, and the corresponding constants are $d_3\zd d_n$, that
is, $f'_i(x,y)=f_i(x,y,d_3,\ldots,d_{n})$. Then the following is a
strict $(n-1)$-implementation of $f'_i$:
$$f'_i(x,y)+(n-2)=\max_{z_3\zd
z_n}[f_i(x,y,z_3,\ldots,z_{n})+u_{\{d_3\}}(z_3)+\ldots
+u_{\{d_n\}}(z_n)].$$
By Lemma~\ref{strict}, it now is sufficient to show the result for
$\F$ consisting of at most three binary predicates. We can assume
that $\F$ is minimal with the property of not being supermodular
on any chain. In addition, we can assume that the binary
predicates in $\F$ do not contain any all-one column or all-one
row (this is justified by Lemma~\ref{all-one}). We need to
consider three cases depending on the number of predicates in
$\F$.

\blankline

\noindent
\underline{Case 1} $|\F|=1$.\\
The result is proved in Lemma~\ref{morehardcases}(1-2).

\blankline

\noindent
\underline{Case 2} $|\F|=2$.\\
Assume $\F=\{g,h\}$. We consider two subcases:

\begin{enumerate}
\item $g$ is supermodular on $C_1$ and $C_2$ but not on $C_0$;
this implies that $h \in \Q_0$ because otherwise $\F\cup \U_D$ is
supermodular on $C_1$ or $C_2$, or else $h$ is not supermodular on
any chain, contradicting the minimality of $\F$. By
Lemma~\ref{all-one}, we can assume that neither $g$ nor $h$ have
an all-1 row or column. It can be easily checked by inspecting the
list of binary predicates (see, e.g., Fig.~\ref{spmodlist}) that
there exist only three such predicates $g$. These are predicates
$h_3,h_4$ and $h_7$ from Fig.~\ref{spmodlist}. We have that
$h_4(x,y)+1=h_3(x,y)+u_{\{0\}}(x)+u_{\{0\}}(y)$ is a strict
2-implementation of $h_4$ from $h_3$,
$h_3(x,y)+1=h_4(x,y)+u_{\{1,2\}}(x)+u_{\{1,2\}}(y)$ is a strict
2-implementation of $h_3$ from $h_4$, and
$h_7(x,y)=h_3(x,y)+h_4(x,y)$ is a strict 1-implementation of
$h_7$. Hence, since all unary predicates are available, it is
enough to show the result for $g=h_7$, which has already been
obtained in Lemma~\ref{morehardcases}(3).

\item None of the predicates $g,h$ is supermodular on two distinct
(that is, not mutually dual) chains. By symmetry, we may assume
that $g \in \Q_1$ and $h \in \Q_0$. Then the result follows from
Lemma~\ref{morehardcases}(4).
\end{enumerate}


\noindent
\underline{Case 3} $|\F|=3$.\\
By the minimality of $\F$, it follows that $\F=\{g_0,g_1,g_2\}$
where each $g_i$ is {\em not} supermodular
on $C_i$, but is supermodular on the other two chains.
As argued in the previous case, we may assume that
$g_0=h_7$. By symmetry, we may assume that
$g_1$ and $g_2$ have the following matrices, respectively:

\[ \begin{array}{ccc} 1 & 0 & 1 \\ 0 & 1 & 0 \\ 1 & 0 & 1 \end{array}
 \; \; {\rm and} \; \;  \begin{array}{ccc} 1 & 1 & 0
\\ 1 & 1 & 0 \\ 0 & 0 & 1 \end{array} .\]

\noindent It remains to say that, for such $\F$,
$\APX$-completeness of $\MCSP{\F\cup\U_D}$ was shown in
Example~\ref{3to1example}.
\end{proof}

\subsection{The Last Step}\label{generalsec}

We will need one more auxiliary lemma.
Let ${\cal C}_D=\{u_{\{d\}} \; | \; d \in D\}$.

\begin{lemma}\label{allunary}
For any $\F$, if $\MCSP{\F\cup \U_D}$ is $\APX$-complete, then so
is $\MCSP{\F\cup \C_D}$.
\end{lemma}
\begin{proof}
For any disjoint subsets $S,T$ of $D$, $u_{S\cup
T}(x)=u_S(x)+u_T(x)$ is a strict 1-implementation of $u_{S\cup
T}$. Use this repeatedly and apply Lemma~\ref{strict}.
\end{proof}

\begin{proposition}\label{general}
If $\F$ is not supermodular on any chain, then $\MCSP\F$ is $\APX$-complete.
\end{proposition}
\begin{proof}
If $\F$ contains a non-trivial irreflexive predicate then the
result follows from Proposition~\ref{irrefl}. Letting $r(f)=\{d
\in D \; | \; f(d,\ldots,d)=1\}$ for a predicate $f$, we can now
assume that $r(f)\ne\emptyset$ for all $f\in\F$. Let $r(\F)=\{r(f)
\; | \; f \in \F\}$. If $r(f)=S$ then  $u_S(x)=f(x\zd x)$ is a
strict 1-implementation of $u_S(x)$. Hence, for each $S \in
r(\F)$, we can without loss of generality assume that $u_S\in\F$.
Note that if, for some $d\in D$, we have $d\in r(f)$ for all $f\in
\F$, then the operation sending all elements of $D$ to $d$ is an
endomorphism of $\F$, contradicting the assumption that $\F$ is a
core. Hence, for every $d\in D$, there is a unary predicate
$u_S\in \F$ (depending on $d$) such that $d\not\in S$.

Note that if $\{a,b,c\}=D$ then
$u_{\{b\}}(x)+1=u_{\{a,b\}}(x)+u_{\{b,c\}}(x)$ is a strict
2-implementation of  $u_{\{b\}}(x)$. Hence, we may assume that,
for any distinct two-element sets $S_1,S_2$ in $r(\F)$, we also
have $S_1\cap S_2\in r(\F)$. It is easy to see that then $r(\F)$
contains at least one of the following: 1) two distinct
singletons, or 2) sets $\{a,b\}$ and $\{c\}$ such that
$\{a,b,c\}=D$. We will consider these two cases separately.

Note that, by Proposition~\ref{withun}, $\MCSP{\F\cup\U_D}$
is \APX-complete. Then, by Lemma~\ref{allunary}, $\MCSP{\F\cup\C_D}$
is \APX-complete as well. Hence, by Lemma~\ref{strict},
showing that $\F$ can strictly implement every predicate in $\C_D$
is sufficient to prove the proposition.

\blankline

\noindent
\underline{Case 1} $u_{\{a\}},u_{\{b\}}\in \F$ and $a\ne b$.\\
Assume without loss of generality that $a=0$ and $b=1$. We will
show that $\F$ can strictly implement $u_{\{2\}}$. Since $\F$ is a
core, let $f_1\in \F$ be an $n$-ary predicate witnessing that the
operation $\pi_1$ such that $\pi_1(0)=0$ and $\pi_1(1)=\pi_1(2)=1$
is not an endomorphism of $\F$. Let $\ba=(a_1\zd a_n)$ be a tuple
such that $f({\bf a})=1$, but $f(\pi_1({\bf a}))=0$ (where
$\pi_1({\bf a})=(\pi_1(a_1)\zd \pi_1(a_n))$). Note that at least
one of the $a_i$'s must be equal to 2, since otherwise ${\bf
a}=\pi_1({\bf a})$. For each $1\le i\le n$, let $t_i$ be $x$ if
$a_i=2$ and $z_i$ otherwise. Denote by $l$ the number of $t_i$'s
that are of the form $z_i$. Now it is not difficult to verify that
$$g_1(x)+l=\max_{\{z_i\mid a_i\ne 2\}}[f_1(t_1\zd t_n)+
\sum_{a_i\ne 2}{u_{\{a_i\}}(z_i)}]$$
is a strict ($l+1$)-implementation of a unary predicate $g_1(x)$
such that $g_1(2)=1$ and $g_1(1)=0$. That is, $g_1$ is either
$u_{\{2\}}$ or $u_{\{0,2\}}$. If $g_1=u_{\{2\}}$ then we have all
predicates from $\C_D$, and we are done. So assume that
$g_1=u_{\{0,2\}}$.

Next, the operation $\pi_2$ such that $\pi_2(1)=1$ and
$\pi_2(0)=\pi_2(2)=0$ is not an endomorphism of $\F$ either.
Acting as above, one can show that $\F$ strictly implements a
unary predicate $g_2(x)$ such that $g(2)=1$ and $g(0)=0$, which is
either $u_{\{2\}}$ or $u_{\{1,2\}}$. Again, if $g_2=u_{\{2\}}$
then we are done. Otherwise, $g_2=u_{\{1,2\}}$ and
$u_{\{2\}}(x)+1=g_1(x)+g_2(x)$ is strict 2-implementation of
$u_{\{2\}}$.

\blankline

\noindent
\underline{Case 2.} $u_{\{a,b\}},u_{\{c\}}\in\F$ and $\{a,b,c\}=D$.\\
Assume without loss of generality that $a=0$, $b=1$ and $c=2$. Let
$f\in \F$ be a predicate witnessing that the operation $\pi$ such
that $\pi(0)=\pi(1)=1$ and $\pi(2)=2$ is not an endomorphism of
$\F$. If $f$ is unary then $f=u_{\{0\}}$ or $f=u_{\{0,2\}}$. In
the former case we go back to Case 1, and in the latter case
$u_{\{0\}}(x)+1=u_{\{0,1\}}(x)+u_{\{0,2\}}(x)$ is a strict
2-implementation of $u_{\{0\}}$, so we can use Case 1 again.

Assume that $f$ is $n$-ary, $n\ge 2$. Similarly to Case 1,
let $\ba=(a_1\zd a_n)$ be a tuple such that
$f({\bf a})=1$, but $f(\pi({\bf a}))=0$.
For each $1\le i\le n$, let $t_i$ be $x$ if $a_i=0$,
$y$ if $a_i=1$ and $z$ otherwise.
Note that $y$ or/and $z$ may not appear among the $t_i$'s (unlike
$x$ which does appear). We consider the case when $z$ does appear,
the other case is very similar. If none of the $t_i$'s is $y$ then
$g_1(x)=\max_{z}[f_1(t_1\zd t_n)+u_{\{2\}}(z)]$ is a strict
2-implementation of a unary predicate $g_1$ which is either
$u_{\{0\}}$ or $u_{\{0,2\}}$ (since $\pi$ is not its
endomorphism). Hence we are done, as above. Assume now that some
$t_i$ is $y$. Now it is not difficult to verify that
$g_2(x,y)=\max_{z}[f_1(t_1\zd t_n)+u_{\{2\}}(z)]$ is a strict
2-implementation of a binary predicate $g_2$ which satisfies
$g_2(0,1)=1$ and $g_2(1,1)=0$. If $g_2(0,0)=1$, then the predicate
$g_2(x,x)$ is either $u_{\{0\}}$ or $u_{\{0,2\}}$, and we are
done. Otherwise, we have $g_2(0,0)=0$. Now apply
Lemma~\ref{morehardcases}(1) if $g_2(2,2)=0$, and use
Lemma~\ref{morehardcases}(5) otherwise.
\end{proof}

\section{Conclusion}\label{conclsec}

We have proved a dichotomy result for maximum constraint
satisfaction problems over a three-element domain. The property of
supermodularity appears to be the dividing line: those sets of
predicates whose cores have this property give rise to problems
solvable exactly in polynomial time, while all other sets of
predicates can implement, in a regular way, the disequality
predicate on a two- or three-element set, and hence give rise to
\APX-complete problems. Interestingly, the description of
polynomial cases is based on orderings of the domain, which is not
suggested in any way by the formulation of the problem.

It can be shown using Theorem~\ref{tract} that Theorem~\ref{main},
as stated in the paper, does not hold for domains with at least
four elements. The reason is that all lattices on at most
three-element set are chains, but on larger sets there are other
types of lattices (for example, a Boolean lattice on a
four-element set). Corollary~1 of \cite{Krokhin04:diamondsTR}
implies the existence of sets $\F$ such that $\MCSP\F$ is
tractable, and $\F$ is supermodular on some distributive lattice
which is not a chain, but not supermodular on any chain. Hence,
more general lattices are required to make further progress in
classifying the complexity of {\sc Max CSP}s, as is a better
understanding of the supermodularity property on arbitrary
lattices. We believe that the ideas from this paper can be further
developed to obtain a complete classification of approximability
of {\sc Max CSP}.

Notably, the hard problems of the form considered in this paper
do not have a PTAS. It is possible that, as it is done
in Theorem 8.8 of \cite{Creignouetal:siam01}, certain restrictions on
the incidence graph of variables in the instance
can give rise to \NP-hard problems that do have a PTAS.

Finally, techniques of~\cite{Trevisan00:gadgets} can be used
to obtain better implementations and more precise (in)approximability
results for non-Boolean problems {\sc Max CSP}. We leave this direction
for future research.

\section*{Acknowledgements}

The authors wish to thank Peter Jeavons and Benoit Larose
for providing detailed comments on this paper.
Peter Jonsson and Mikael Klasson were supported by the {\em Swedish
Research Council} (VR) under grant 621--2003--3421, and Peter
Jonsson was also supported by the {\em Center for Industrial
Information Technology} (CENIIT) under grant 04.01.
Andrei Krokhin was supported by UK EPSRC grant GR/T05325/01.

\newpage

\section*{Appendix}

\subsection*{Proof of Lemma~\ref{morehardcases} (Cases 2-5)}

In each case, we generate a list of all applicable predicates,
and then optimise it as follows:


\begin{itemize}
    \item skip a predicate $f(x,y)$ if $f'(x,y)$ is in the list, with $f(x,y)=f'(y,x)$;
    \item skip all predicates with an all-1 row or column.

\end{itemize}

By Lemma~\ref{all-one}, it is sufficient to prove the result
for the optimised lists.

Cases 2-5 follow in order, with a description and a list of strict
implementations.
Each strict implementation produces either $neq_3$, or some binary
predicate whose core is a non-supermodular predicate on a two-element
subset of $D$, or some other predicate for which an implementation has
already been found.

Cases 2 and 3 will also use $eq_3$ -- the binary equality
predicate -- as implementation target. Note that it was shown in
Example~\ref{neqfromeq} that the set $\{eq_3\}\cup \U_D$ can
strictly implement $neq_3$, and hence, $\MCSP{\{eq_3\}\cup \U_D}$
is \APX-complete.

\medskip

If some implementation produces a predicate $g$, whose core is
a non-supermodular predicate on a two-element subset of $D$,
then we write $([0,1,2]\mapsto [\pi(0),\pi(1),\pi(2)])$ to
describe the endomorphism $\pi$  leading to the core,
and we also give a matrix for that non-supermodular predicate.

In all strict implementations in this section,
the variables $x,y$ are primary, and $z,w$ (when they appear)
are auxiliary.


\noindent
\subsubsection*{Case 2}
${\cal F}=\{f\} \cup {\mathcal U}_D$ and $f$
is not supermodular on any chain on $D$.

\medskip

We further optimise the list of predicates for this case. We can
assume that $f(d,d)=1$ for some $d$, as the other predicates are
handled in Case 1. By symmetry, we may skip a predicate if there
is another predicate in the list by renaming the elements of $D$.
We can also skip $eq_3$, as we have handled it separately in
Example~\ref{neqfromeq}.


{\flushleft
{\small $\left\{f_{1} := {\footnotesize \begin{array}{l}
110\\
010\\
000\\
\end{array}
}\right\} \cup$ ${\cal U}_D \stackrel{s\phantom{8pt}}{\Longrightarrow_2}$ ${\footnotesize 
\begin{array}{l}
100\\
010\\
001\\
\end{array}
} =: g\qquad$
$g = eq_3$}
\\
{\small Implementation: $g(x,y) + 1 = f_{1}(x,y)+f_{1}(y,x)+u_{\{2\}}(x)+u_{\{2\}}(y)$}

\vskip 0.3cm

{\small $\left\{f_{2} := {\footnotesize \begin{array}{l}
101\\
101\\
000\\
\end{array}
}\right\} \cup$ ${\cal U}_D \stackrel{s\phantom{8pt}}{\Longrightarrow_2}$ ${\footnotesize 
\begin{array}{l}
000\\
000\\
010\\
\end{array}
} =: g\qquad$
$g$ has core $\footnotesize \begin{array}{l}
00\\
10\\
\end{array}
([0,1,2] \mapsto [0,0,1])$}
\\
{\small Implementation: $g(x,y) + 1 = f_{2}(x,y)+u_{\{2\}}(x)+u_{\{1\}}(y)$}

\vskip 0.3cm

{\small $\left\{f_{3} := {\footnotesize \begin{array}{l}
110\\
001\\
100\\
\end{array}
}\right\} \cup$ ${\cal U}_D \stackrel{s\phantom{8pt}}{\Longrightarrow_3}$ ${\footnotesize 
\begin{array}{l}
000\\
001\\
010\\
\end{array}
} =: g\qquad$
$g$ has core $\footnotesize \begin{array}{l}
01\\
10\\
\end{array}
([0,1,2] \mapsto [0,1,0])$}
\\
{\small Implementation: $g(x,y) + 2 = f_{3}(x,y)+f_{3}(y,x)+u_{\{1,2\}}(x)+u_{\{1,2\}}(y)$}

\vskip 0.3cm

{\small $\left\{f_{4} := {\footnotesize \begin{array}{l}
110\\
011\\
101\\
\end{array}
}\right\} \cup$ ${\cal U}_D \stackrel{s\phantom{8pt}}{\Longrightarrow_2}$ ${\footnotesize 
\begin{array}{l}
100\\
010\\
001\\
\end{array}
} =: g\qquad$
$g = eq_3$}
\\
{\small Implementation: $g(x,y) + 1 = f_{4}(x,y)+f_{4}(y,x)$}

\vskip 0.3cm

{\small $\left\{f_{5} := {\footnotesize \begin{array}{l}
101\\
100\\
000\\
\end{array}
}\right\} \cup$ ${\cal U}_D \stackrel{s\phantom{8pt}}{\Longrightarrow_3}$ ${\footnotesize 
\begin{array}{l}
000\\
101\\
000\\
\end{array}
} =: g\qquad$
$g$ has core $\footnotesize \begin{array}{l}
00\\
10\\
\end{array}
([0,1,2] \mapsto [0,1,0])$}
\\
{\small Implementation: $g(x,y) + 2 = max_{z}[f_{5}(z,y)+f_{5}(x,z)+u_{\{2\}}(z)+u_{\{1,2\}}(x)]$}

\vskip 0.3cm

{\small $\left\{f_{6} := {\footnotesize \begin{array}{l}
001\\
010\\
000\\
\end{array}
}\right\} \cup$ ${\cal U}_D \stackrel{s\phantom{8pt}}{\Longrightarrow_2}$ ${\footnotesize 
\begin{array}{l}
110\\
010\\
000\\
\end{array}
} =: g\qquad$
$g = f_{1}$}
\\
{\small Implementation: $g(x,y) + 1 = max_{z}[f_{6}(z,x)+f_{6}(y,z)+u_{\{0\}}(x)]$}

\vskip 0.3cm

{\small $\left\{f_{7} := {\footnotesize \begin{array}{l}
011\\
010\\
000\\
\end{array}
}\right\} \cup$ ${\cal U}_D \stackrel{s\phantom{8pt}}{\Longrightarrow_3}$ ${\footnotesize 
\begin{array}{l}
110\\
010\\
000\\
\end{array}
} =: g\qquad$
$g = f_{1}$}
\\
{\small Implementation: $g(x,y) + 2 = max_{z}[f_{7}(z,x)+f_{7}(z,y)+f_{7}(y,z)+u_{\{2\}}(z)+u_{\{0\}}(x)]$}

\vskip 0.3cm

{\small $\left\{f_{8} := {\footnotesize \begin{array}{l}
001\\
110\\
000\\
\end{array}
}\right\} \cup$ ${\cal U}_D \stackrel{s\phantom{8pt}}{\Longrightarrow_4}$ ${\footnotesize 
\begin{array}{l}
001\\
000\\
000\\
\end{array}
} =: g\qquad$
$g$ has core $\footnotesize \begin{array}{l}
01\\
00\\
\end{array}
([0,1,2] \mapsto [0,0,1])$}
\\
{\small Implementation: $g(x,y) + 3 = max_{z}[f_{8}(z,y)+f_{8}(x,z)+u_{\{1,2\}}(z)+u_{\{0,2\}}(x)+u_{\{2\}}(y)]$}

\vskip 0.3cm

{\small $\left\{f_{9} := {\footnotesize \begin{array}{l}
101\\
110\\
000\\
\end{array}
}\right\} \cup$ ${\cal U}_D \stackrel{s\phantom{8pt}}{\Longrightarrow_4}$ ${\footnotesize 
\begin{array}{l}
001\\
000\\
000\\
\end{array}
} =: g\qquad$
$g$ has core $\footnotesize \begin{array}{l}
01\\
00\\
\end{array}
([0,1,2] \mapsto [0,0,1])$}
\\
{\small Implementation: $g(x,y) + 3 = max_{z}[f_{9}(z,y)+f_{9}(x,z)+u_{\{1\}}(z)+u_{\{0,2\}}(x)+u_{\{2\}}(y)]$}

\vskip 0.3cm

{\small $\left\{f_{10} := {\footnotesize \begin{array}{l}
011\\
110\\
000\\
\end{array}
}\right\} \cup$ ${\cal U}_D \stackrel{s\phantom{8pt}}{\Longrightarrow_4}$ ${\footnotesize 
\begin{array}{l}
001\\
010\\
000\\
\end{array}
} =: g\qquad$
$g = f_{6}$}
\\
{\small Implementation: $g(x,y) + 3 = max_{z}[f_{10}(x,z)+f_{10}(x,y)+f_{10}(y,z)+u_{\{0,2\}}(z)+u_{\{2\}}(x)+u_{\{2\}}(y)]$}

\vskip 0.3cm

{\small $\left\{f_{11} := {\footnotesize \begin{array}{l}
001\\
010\\
100\\
\end{array}
}\right\} \cup$ ${\cal U}_D \stackrel{s\phantom{8pt}}{\Longrightarrow_2}$ ${\footnotesize 
\begin{array}{l}
100\\
010\\
001\\
\end{array}
} =: g\qquad$
$g = eq_3$}
\\
{\small Implementation: $g(x,y) + 1 = max_{z}[f_{11}(z,x)+f_{11}(z,y)]$}

\vskip 0.3cm

{\small $\left\{f_{12} := {\footnotesize \begin{array}{l}
011\\
010\\
100\\
\end{array}
}\right\} \cup$ ${\cal U}_D \stackrel{s\phantom{8pt}}{\Longrightarrow_3}$ ${\footnotesize 
\begin{array}{l}
110\\
010\\
000\\
\end{array}
} =: g\qquad$
$g = f_{1}$}
\\
{\small Implementation: $g(x,y) + 2 = max_{z}[f_{12}(z,y)+f_{12}(x,z)+u_{\{1,2\}}(z)]$}

\vskip 0.3cm

{\small $\left\{f_{13} := {\footnotesize \begin{array}{l}
011\\
110\\
100\\
\end{array}
}\right\} \cup$ ${\cal U}_D \stackrel{s\phantom{8pt}}{\Longrightarrow_4}$ ${\footnotesize 
\begin{array}{l}
011\\
010\\
000\\
\end{array}
} =: g\qquad$
$g = f_{7}$}
\\
{\small Implementation: $g(x,y) + 3 = max_{z}[f_{13}(z,x)+f_{13}(z,y)+f_{13}(x,y)+u_{\{0\}}(z)+u_{\{0\}}(x)]$}

\vskip 0.3cm

{\small $\left\{f_{14} := {\footnotesize \begin{array}{l}
110\\
011\\
100\\
\end{array}
}\right\} \cup$ ${\cal U}_D \stackrel{s\phantom{8pt}}{\Longrightarrow_3}$ ${\footnotesize 
\begin{array}{l}
000\\
000\\
110\\
\end{array}
} =: g\qquad$
$g$ has core $\footnotesize \begin{array}{l}
00\\
10\\
\end{array}
([0,1,2] \mapsto [0,0,1])$}
\\
{\small Implementation: $g(x,y) + 2 = max_{z}[f_{14}(z,y)+f_{14}(x,z)+u_{\{2\}}(x)]$}

\vskip 0.3cm

{\small $\left\{f_{15} := {\footnotesize \begin{array}{l}
101\\
011\\
100\\
\end{array}
}\right\} \cup$ ${\cal U}_D \stackrel{s\phantom{8pt}}{\Longrightarrow_4}$ ${\footnotesize 
\begin{array}{l}
110\\
010\\
000\\
\end{array}
} =: g\qquad$
$g = f_{1}$}
\\
{\small Implementation: $g(x,y) + 3 = max_{z}[f_{15}(z,x)+f_{15}(z,y)+f_{15}(x,z)+u_{\{2\}}(z)+u_{\{1\}}(y)]$}

\vskip 0.3cm

{\small $\left\{f_{16} := {\footnotesize \begin{array}{l}
011\\
011\\
100\\
\end{array}
}\right\} \cup$ ${\cal U}_D \stackrel{s\phantom{8pt}}{\Longrightarrow_4}$ ${\footnotesize 
\begin{array}{l}
000\\
000\\
100\\
\end{array}
} =: g\qquad$
$g$ has core $\footnotesize \begin{array}{l}
00\\
10\\
\end{array}
([0,1,2] \mapsto [0,0,1])$}
\\
{\small Implementation: $g(x,y) + 3 = max_{z}[f_{16}(z,y)+f_{16}(x,z)+f_{16}(x,y)+u_{\{2\}}(z)+u_{\{2\}}(x)]$}

\vskip 0.3cm

{\small $\left\{f_{17} := {\footnotesize \begin{array}{l}
110\\
010\\
001\\
\end{array}
}\right\} \cup$ ${\cal U}_D \stackrel{s\phantom{8pt}}{\Longrightarrow_3}$ ${\footnotesize 
\begin{array}{l}
110\\
010\\
000\\
\end{array}
} =: g\qquad$
$g = f_{1}$}
\\
{\small Implementation: $g(x,y) + 2 = max_{z}[f_{17}(z,y)+f_{17}(x,z)+u_{\{0,1\}}(z)]$}

\vskip 0.3cm

{\small $\left\{f_{18} := {\footnotesize \begin{array}{l}
101\\
110\\
001\\
\end{array}
}\right\} \cup$ ${\cal U}_D \stackrel{s\phantom{8pt}}{\Longrightarrow_3}$ ${\footnotesize 
\begin{array}{l}
110\\
010\\
000\\
\end{array}
} =: g\qquad$
$g = f_{1}$}
\\
{\small Implementation: $g(x,y) + 2 = max_{z}[f_{18}(z,x)+f_{18}(y,z)+u_{\{0,1\}}(x)]$}

\vskip 0.3cm

{\small $\left\{f_{19} := {\footnotesize \begin{array}{l}
110\\
100\\
000\\
\end{array}
}\right\} \cup$ ${\cal U}_D \stackrel{s\phantom{8pt}}{\Longrightarrow_5}$ ${\footnotesize 
\begin{array}{l}
000\\
101\\
000\\
\end{array}
} =: g\qquad$
$g$ has core $\footnotesize \begin{array}{l}
00\\
10\\
\end{array}
([0,1,2] \mapsto [0,1,0])$}
\\
{\small Implementation: $g(x,y) + 4 = max_{z,w}[f_{19}(z,w)+f_{19}(z,y)+f_{19}(w,x)+u_{\{1\}}(z)+u_{\{1\}}(w)+u_{\{1,2\}}(x)+u_{\{2\}}(y)]$}

\vskip 0.3cm

{\small $\left\{f_{20} := {\footnotesize \begin{array}{l}
101\\
010\\
100\\
\end{array}
}\right\} \cup$ ${\cal U}_D \stackrel{s\phantom{8pt}}{\Longrightarrow_5}$ ${\footnotesize 
\begin{array}{l}
001\\
010\\
000\\
\end{array}
} =: g\qquad$
$g = f_{6}$}
\\
{\small Implementation: $g(x,y) + 4 = max_{z,w}[f_{20}(z,w)+f_{20}(z,y)+f_{20}(w,x)+u_{\{2\}}(z)+u_{\{1,2\}}(w)+u_{\{1,2\}}(y)]$}

\vskip 0.3cm

{\small $\left\{f_{21} := {\footnotesize \begin{array}{l}
101\\
011\\
110\\
\end{array}
}\right\} \cup$ ${\cal U}_D \stackrel{s\phantom{8pt}}{\Longrightarrow_4}$ ${\footnotesize 
\begin{array}{l}
101\\
010\\
100\\
\end{array}
} =: g\qquad$
$g = f_{20}$}
\\
{\small Implementation: $g(x,y) + 3 = max_{z}[f_{21}(z,x)+f_{21}(z,y)+f_{21}(x,y)+u_{\{0,2\}}(z)]$}


}


\bigskip

\subsubsection*{Case 3}
${\cal F}=\{f,h_7\} \cup {\mathcal U}_D$ and $f \in \Q_0$.

\medskip

We further optimise the list of predicates for this case as
follows. By symmetry, we can skip a predicate if there is another
predicate in the list obtained by swapping the names of elements
$1$ and $2$ of $D$.


{\flushleft
{\small $\left\{f_{1} := {\footnotesize \begin{array}{l}
000\\
000\\
101\\
\end{array}
},h_7\right\} \cup$ ${\cal U}_D \stackrel{s\phantom{8pt}}{\Longrightarrow_3}$ ${\footnotesize 
\begin{array}{l}
100\\
010\\
001\\
\end{array}
} =: g\qquad$
$g = eq_3$
\\Implementation: $g(x,y) + 2 = f_{1}(x,y)+f_{1}(y,x)+h_7(x,y)+u_{\{0,1\}}(x)+u_{\{0,1\}}(y)$\\
}\vskip 0.3cm
{\small $\left\{f_{2} := {\footnotesize \begin{array}{l}
000\\
110\\
101\\
\end{array}
},h_7\right\} \cup$ ${\cal U}_D \stackrel{s\phantom{8pt}}{\Longrightarrow_2}$ ${\footnotesize 
\begin{array}{l}
100\\
010\\
001\\
\end{array}
} =: g\qquad$
$g = eq_3$
\\Implementation: $g(x,y) + 1 = f_{2}(x,y)+h_7(x,y)+u_{\{0\}}(x)$\\
}\vskip 0.3cm
{\small $\left\{f_{3} := {\footnotesize \begin{array}{l}
000\\
010\\
001\\
\end{array}
}\right\} \cup$ ${\cal U}_D \stackrel{s\phantom{8pt}}{\Longrightarrow_2}$ ${\footnotesize 
\begin{array}{l}
000\\
110\\
101\\
\end{array}
} =: g\qquad$
$g = f_{2}$
\\Implementation: $g(x,y) + 1 = max_{z}[f_{3}(z,x)+f_{3}(z,y)+u_{\{0\}}(y)]$\\
}\vskip 0.3cm
{\small $\left\{f_{4} := {\footnotesize \begin{array}{l}
001\\
110\\
001\\
\end{array}
}\right\} \cup$ ${\cal U}_D \stackrel{s\phantom{8pt}}{\Longrightarrow_3}$ ${\footnotesize 
\begin{array}{l}
000\\
000\\
101\\
\end{array}
} =: g\qquad$
$g = f_{1}$
\\Implementation: $g(x,y) + 2 = max_{z}[f_{4}(z,x)+f_{4}(y,z)+u_{\{2\}}(z)]$\\
}\vskip 0.3cm
{\small $\left\{f_{5} := {\footnotesize \begin{array}{l}
000\\
010\\
101\\
\end{array}
}\right\} \cup$ ${\cal U}_D \stackrel{s\phantom{8pt}}{\Longrightarrow_2}$ ${\footnotesize 
\begin{array}{l}
000\\
110\\
101\\
\end{array}
} =: g\qquad$
$g = f_{2}$
\\Implementation: $g(x,y) + 1 = max_{z}[f_{5}(x,z)+f_{5}(y,z)+u_{\{0\}}(y)]$\\
}

}


\subsection*{Case 4}
${\cal F}=\{f,h\} \cup {\cal U}_D$ and $f \in \Q_1$ and $h \in
\Q_0$.

\medskip

We show that $\{f\} \cup {\cal U}_D$ can strictly implement $h_7$,
whereby the result follows from Case 3.


{\flushleft
{\small $\left\{f_{1} := {\footnotesize \begin{array}{l}
100\\
000\\
011\\
\end{array}
}\right\} \cup$ ${\cal U}_D \stackrel{s\phantom{8pt}}{\Longrightarrow_2}$ ${\footnotesize 
\begin{array}{l}
100\\
011\\
011\\
\end{array}
} =: g\qquad$
$g = h_7$
\\Implementation: $g(x,y) + 1 = max_{z}[f_{1}(z,x)+f_{1}(z,y)]$\\
}\vskip 0.3cm
{\small $\left\{f_{2} := {\footnotesize \begin{array}{l}
100\\
100\\
011\\
\end{array}
}\right\} \cup$ ${\cal U}_D \stackrel{s\phantom{8pt}}{\Longrightarrow_2}$ ${\footnotesize 
\begin{array}{l}
100\\
011\\
011\\
\end{array}
} =: g\qquad$
$g = h_7$
\\Implementation: $g(x,y) + 1 = max_{z}[f_{2}(z,x)+f_{2}(z,y)]$\\
}\vskip 0.3cm
{\small $\left\{f_{3} := {\footnotesize \begin{array}{l}
100\\
100\\
000\\
\end{array}
}\right\} \cup$ ${\cal U}_D \stackrel{s\phantom{8pt}}{\Longrightarrow_3}$ ${\footnotesize 
\begin{array}{l}
100\\
000\\
011\\
\end{array}
} =: g\qquad$
$g = f_{1}$
\\Implementation: $g(x,y) + 2 = max_{z}[f_{3}(z,x)+f_{3}(z,y)+f_{3}(x,y)+u_{\{2\}}(z)+u_{\{2\}}(x)+u_{\{1,2\}}(y)]$\\
}\vskip 0.3cm
{\small $\left\{f_{4} := {\footnotesize \begin{array}{l}
100\\
100\\
001\\
\end{array}
}\right\} \cup$ ${\cal U}_D \stackrel{s\phantom{8pt}}{\Longrightarrow_4}$ ${\footnotesize 
\begin{array}{l}
100\\
100\\
000\\
\end{array}
} =: g\qquad$
$g = f_{3}$
\\Implementation: $g(x,y) + 3 = max_{z}[f_{4}(z,y)+f_{4}(x,z)+u_{\{2\}}(z)+u_{\{0,1\}}(x)+u_{\{0,1\}}(y)]$\\
}\vskip 0.3cm
{\small $\left\{f_{5} := {\footnotesize \begin{array}{l}
100\\
101\\
001\\
\end{array}
}\right\} \cup$ ${\cal U}_D \stackrel{s\phantom{8pt}}{\Longrightarrow_4}$ ${\footnotesize 
\begin{array}{l}
100\\
100\\
001\\
\end{array}
} =: g\qquad$
$g = f_{4}$
\\Implementation: $g(x,y) + 3 = max_{z}[f_{5}(z,x)+f_{5}(z,y)+f_{5}(x,z)+u_{\{0\}}(z)+u_{\{1,2\}}(x)]$\\
}\vskip 0.2cm
{\small $\left\{f_{6} := {\footnotesize \begin{array}{l}
000\\
000\\
011\\
\end{array}
}\right\} \cup$ ${\cal U}_D \stackrel{s\phantom{8pt}}{\Longrightarrow_2}$ ${\footnotesize 
\begin{array}{l}
100\\
100\\
000\\
\end{array}
} =: g\qquad$
$g = f_{3}$
\\Implementation: $g(x,y) + 1 = f_{6}(x,y)+u_{\{0,1\}}(x)+u_{\{0\}}(y)$\\
}\vskip 0.2cm
{\small $\left\{f_{7} := {\footnotesize \begin{array}{l}
100\\
000\\
001\\
\end{array}
}\right\} \cup$ ${\cal U}_D \stackrel{s\phantom{8pt}}{\Longrightarrow_2}$ ${\footnotesize 
\begin{array}{l}
100\\
101\\
001\\
\end{array}
} =: g\qquad$
$g = f_{5}$
\\Implementation: $g(x,y) + 1 = max_{z}[f_{7}(z,x)+f_{7}(z,y)+u_{\{1\}}(x)]$\\
}

}


\subsection*{Case 5}
${\cal F}=\{f,u_{\{0,1\}}\}$ where $f$ is such that $f(0,0) = f(1,1) = 0$ and $f(2,2) = f(0,1) = 1$.


{\flushleft
{\small $\left\{f_{1} := {\footnotesize \begin{array}{l}
010\\
001\\
101\\
\end{array}
},u_{\{0,1\}}\right\} \stackrel{s\phantom{8pt}}{\Longrightarrow_3}$ ${\footnotesize 
\begin{array}{l}
010\\
100\\
000\\
\end{array}
} =: g\qquad$
$g$ has core $\footnotesize \begin{array}{l}
01\\
10\\
\end{array}
([0,1,2] \mapsto [1,0,0])$
\\Implementation: $g(x,y) + 2 = f_{1}(x,y)+f_{1}(y,x)+u_{\{0,1\}}(x)+u_{\{0,1\}}(y)$\\
}\vskip 0.3cm
{\small $\left\{f_{2} := {\footnotesize \begin{array}{l}
011\\
000\\
011\\
\end{array}
},u_{\{0,1\}}\right\} \stackrel{s\phantom{8pt}}{\Longrightarrow_3}$ ${\footnotesize 
\begin{array}{l}
010\\
100\\
000\\
\end{array}
} =: g\qquad$
$g$ has core $\footnotesize \begin{array}{l}
01\\
10\\
\end{array}
([0,1,2] \mapsto [1,0,0])$
\\Implementation: $g(x,y) + 2 = f_{2}(x,y)+f_{2}(y,x)+u_{\{0,1\}}(x)+u_{\{0,1\}}(y)$\\
}\vskip 0.3cm
{\small $f_{3} := {\footnotesize \begin{array}{l}
010\\
101\\
101\\
\end{array}
} \stackrel{s\phantom{8pt}}{\Longrightarrow_3}$ ${\footnotesize 
\begin{array}{l}
010\\
001\\
101\\
\end{array}
} =: g\qquad$
$g = f_{1}$
\\Implementation: $g(x,y) + 2 = max_{z}[f_{3}(z,x)+f_{3}(x,y)+f_{3}(y,z)]$\\
}\vskip 0.3cm
{\small $\left\{f_{4} := {\footnotesize \begin{array}{l}
010\\
000\\
001\\
\end{array}
},u_{\{0,1\}}\right\} \stackrel{s\phantom{8pt}}{\Longrightarrow_4}$ ${\footnotesize 
\begin{array}{l}
010\\
000\\
000\\
\end{array}
} =: g\qquad$
$g$ has core $\footnotesize \begin{array}{l}
01\\
00\\
\end{array}
([0,1,2] \mapsto [0,1,0])$
\\Implementation: $g(x,y) + 3 = max_{z,w}[f_{4}(z,w)+f_{4}(w,y)+f_{4}(x,z)+u_{\{0,1\}}(z)+u_{\{0,1\}}(w)]$\\
}\vskip 0.3cm
{\small $\left\{f_{5} := {\footnotesize \begin{array}{l}
011\\
000\\
001\\
\end{array}
},u_{\{0,1\}}\right\} \stackrel{s\phantom{8pt}}{\Longrightarrow_3}$ ${\footnotesize 
\begin{array}{l}
011\\
000\\
011\\
\end{array}
} =: g\qquad$
$g = f_{2}$
\\Implementation: $g(x,y) + 2 = max_{z,w}[f_{5}(z,y)+f_{5}(w,z)+f_{5}(x,w)+u_{\{0,1\}}(z)]$\\
}\vskip 0.3cm
{\small $\left\{f_{6} := {\footnotesize \begin{array}{l}
010\\
100\\
001\\
\end{array}
},u_{\{0,1\}}\right\} \stackrel{s\phantom{8pt}}{\Longrightarrow_4}$ ${\footnotesize 
\begin{array}{l}
010\\
100\\
000\\
\end{array}
} =: g\qquad$
$g$ has core $\footnotesize \begin{array}{l}
01\\
10\\
\end{array}
([0,1,2] \mapsto [1,0,0])$
\\Implementation: $g(x,y) + 3 = max_{z,w}[f_{6}(z,w)+f_{6}(z,y)+f_{6}(w,x)+u_{\{0,1\}}(z)]$\\
}\vskip 0.3cm
{\small $f_{7} := {\footnotesize \begin{array}{l}
011\\
100\\
001\\
\end{array}
} \stackrel{s\phantom{8pt}}{\Longrightarrow_3}$ ${\footnotesize 
\begin{array}{l}
010\\
101\\
101\\
\end{array}
} =: g\qquad$
$g = f_{3}$
\\Implementation: $g(x,y) + 2 = max_{z,w}[f_{7}(w,z)+f_{7}(w,x)+f_{7}(y,z)]$\\
}\vskip 0.3cm
{\small $\left\{f_{8} := {\footnotesize \begin{array}{l}
010\\
001\\
001\\
\end{array}
},u_{\{0,1\}}\right\} \stackrel{s\phantom{8pt}}{\Longrightarrow_4}$ ${\footnotesize 
\begin{array}{l}
000\\
100\\
100\\
\end{array}
} =: g\qquad$
$g$ has core $\footnotesize \begin{array}{l}
00\\
10\\
\end{array}
([0,1,2] \mapsto [0,1,1])$
\\Implementation: $g(x,y) + 3 = max_{z,w}[f_{8}(z,w)+f_{8}(x,w)+f_{8}(y,z)+u_{\{0,1\}}(z)]$\\
}\vskip 0.3cm
{\small $f_{9} := {\footnotesize \begin{array}{l}
010\\
101\\
001\\
\end{array}
} \stackrel{s\phantom{8pt}}{\Longrightarrow_3}$ ${\footnotesize 
\begin{array}{l}
010\\
101\\
101\\
\end{array}
} =: g\qquad$
$g = f_{3}$
\\Implementation: $g(x,y) + 2 = max_{z,w}[f_{9}(w,z)+f_{9}(w,y)+f_{9}(x,z)]$\\
}\vskip 0.3cm
{\small $\left\{f_{10} := {\footnotesize \begin{array}{l}
010\\
000\\
101\\
\end{array}
},u_{\{0,1\}}\right\} \stackrel{s\phantom{8pt}}{\Longrightarrow_4}$ ${\footnotesize 
\begin{array}{l}
010\\
000\\
010\\
\end{array}
} =: g\qquad$
$g$ has core $\footnotesize \begin{array}{l}
01\\
00\\
\end{array}
([0,1,2] \mapsto [0,1,0])$
\\Implementation: $g(x,y) + 3 = max_{z,w}[f_{10}(z,y)+f_{10}(w,z)+f_{10}(w,x)+u_{\{0,1\}}(z)]$\\
}\vskip 0.3cm
{\small $\left\{f_{11} := {\footnotesize \begin{array}{l}
011\\
000\\
101\\
\end{array}
},u_{\{0,1\}}\right\} \stackrel{s\phantom{8pt}}{\Longrightarrow_4}$ ${\footnotesize 
\begin{array}{l}
011\\
000\\
011\\
\end{array}
} =: g\qquad$
$g = f_{2}$
\\Implementation: $g(x,y) + 3 = max_{z,w}[f_{11}(z,w)+f_{11}(z,y)+f_{11}(w,x)+u_{\{0,1\}}(z)]$\\
}\vskip 0.3cm
{\small $\left\{f_{12} := {\footnotesize \begin{array}{l}
011\\
100\\
101\\
\end{array}
},u_{\{0,1\}}\right\} \stackrel{s\phantom{8pt}}{\Longrightarrow_5}$ ${\footnotesize 
\begin{array}{l}
011\\
100\\
100\\
\end{array}
} =: g\qquad$
$g$ has core $\footnotesize \begin{array}{l}
01\\
10\\
\end{array}
([0,1,2] \mapsto [1,0,0])$
\\Implementation: $g(x,y) + 4 = max_{z,w}[f_{12}(z,w)+f_{12}(z,y)+f_{12}(w,x)+u_{\{0,1\}}(z)+u_{\{0,1\}}(w)]$\\
}\vskip 0.3cm
{\small $\left\{f_{13} := {\footnotesize \begin{array}{l}
010\\
000\\
011\\
\end{array}
},u_{\{0,1\}}\right\} \stackrel{s\phantom{8pt}}{\Longrightarrow_3}$ ${\footnotesize 
\begin{array}{l}
011\\
000\\
011\\
\end{array}
} =: g\qquad$
$g = f_{2}$
\\Implementation: $g(x,y) + 2 = max_{z,w}[f_{13}(z,w)+f_{13}(w,y)+f_{13}(x,z)+u_{\{0,1\}}(z)]$\\
}\vskip 0.3cm
{\small $\left\{f_{14} := {\footnotesize \begin{array}{l}
010\\
001\\
011\\
\end{array}
},u_{\{0,1\}}\right\} \stackrel{s\phantom{8pt}}{\Longrightarrow_4}$ ${\footnotesize 
\begin{array}{l}
011\\
000\\
011\\
\end{array}
} =: g\qquad$
$g = f_{2}$
\\Implementation: $g(x,y) + 3 = max_{z,w}[f_{14}(z,w)+f_{14}(x,z)+f_{14}(y,w)+u_{\{0,1\}}(z)]$\\
}\vskip 0.3cm
{\small $\left\{f_{15} := {\footnotesize \begin{array}{l}
010\\
101\\
011\\
\end{array}
},u_{\{0,1\}}\right\} \stackrel{s\phantom{8pt}}{\Longrightarrow_5}$ ${\footnotesize 
\begin{array}{l}
010\\
101\\
010\\
\end{array}
} =: g\qquad$
$g$ has core $\footnotesize \begin{array}{l}
01\\
10\\
\end{array}
([0,1,2] \mapsto [0,1,0])$
\\Implementation: $g(x,y) + 4 = max_{z,w}[f_{15}(z,w)+f_{15}(z,y)+f_{15}(w,x)+u_{\{0,1\}}(z)+u_{\{0,1\}}(w)]$\\
}

}

\end{document}